\documentclass[preprint,compress,12pt]{elsarticle}
\usepackage{bm}
\usepackage{mathtools} 
\usepackage{booktabs}
\usepackage{epsfig}
\usepackage{placeins}
\usepackage{subfig}
\usepackage{graphicx}

\usepackage[latin9]{inputenc}
\usepackage{float}
\usepackage{amsmath}
\usepackage{graphicx}
\usepackage{esint}
\usepackage{array}
\usepackage{float}
\usepackage{amsmath}
\usepackage{stackrel}
\usepackage[version=4]{mhchem}
\usepackage[running]{lineno}

\usepackage{listings}
\usepackage{color}
\usepackage{xcolor}
\usepackage{enumerate}   
\usepackage{soul}
\usepackage[hyphens]{url}  
\usepackage[pdftex,bookmarksnumbered,hidelinks,breaklinks]{hyperref}

\definecolor{dkgreen}{rgb}{0,0.6,0}
\definecolor{gray}{rgb}{0.5,0.5,0.5}
\definecolor{mauve}{rgb}{0.58,0,0.84}

\usepackage{amssymb}
\journal{Journal of Computer Physics Communications}
\begin{document}
	
	\begin{frontmatter}
		
		\title{Parallelization of Garfield++ and neBEM to simulate space-charge effects in RPCs}
			\author[a,b]{Tanay Dey\corref{mycorrespondingauthor}}
		\cortext[mycorrespondingauthor]{Corresponding author}
		\ead{tanay.jop@gmail.com}
		\author[c]{Purba Bhattacharya}
		\author[a,d]{Supratik Mukhopadhyay}
		\author[a,d]{Nayana Majumdar}
		\author[e]{Abhishek Seal}
		\author[b]{Subhasis Chattopadhyay}

			\affiliation[a]{organization={Homi Bhabha National Institute},
			city={Mumbai},
			country={India}}
		\affiliation[b]{organization={Variable Energy Cyclotron centre},
			city={Kolkata},
			country={India}}
		
			\affiliation[c]{organisation={Department of Physics, School of Basic and Applied Sciences, Adamas University}, 
			city={Kolkata}, 
			country={India}}
		
		\affiliation[d]{organization={Saha Institute of Nuclear Physics},
			city={Kolkata},
			country={India}
		}
	\affiliation[e]{Regent Education and Research Foundation, city={Kolkata}, country={India}}
		\begin{abstract}
Numerical simulation of avalanches, saturated avalanches, and streamers can help us understand the detector physics of Resistive Plate Chambers (RPC).
3D Monte Carlo simulation of an avalanche inside an RPC, the transition from avalanche to saturated avalanche to streamer may help the search for the optimum voltage and alternate gas mixtures.
This task is dauntingly resource hungry, especially when space-charge effects become important, which often coincides with important regimes of operation of these devices.
By modifying the electric field inside the RPC dynamically, the space charge plays a crucial role in determining the response of the detector.
In this work, a numerical model has been proposed to calculate the dynamic space-charge field inside an RPC and the same has been implemented in the Garfield++ framework.
By modeling space charge as large number of line charges and using the multithreading technique OpenMP to calculate electric field, drift line, electron gain, and space-charge field, it has been possible to maintain time consumption within reasonable limits.
For this purpose, a new class, pAvalancheMC has been introduced in Garfield++.
The calculations have been successfully verified with those from existing solvers and an example is provided to show the performance of pAvalancheMC.
Moreover, the details of the transition of an avalanche into a saturated avalanche have been discussed. The induced-charge distribution is calculated for a timing RPC and results are verified with the experiment. 
	
		\end{abstract}

		\begin{keyword}
Avalanche-induced secondary effects \sep Charge transport \sep Detector modelling and simulations \sep Electric fields \sep Multiplication and induction \sep OpenMP \sep Parallel computation \sep Resistive-plate chambers \sep Space charge \sep Physics of Gases, Plasmas and Electric Discharges.
		\end{keyword}
		
	\end{frontmatter}
	
	
	\section{Introduction}\label{sec:1_intro}
	Development of gaseous ionization detectors like Gas Electron Multipliers (GEM)\cite{SAULI20162}, Resistive Plate Chambers (RPC)\cite{cardeli-1,cardeli-2}, Time Projection Chambers (TPC)\cite{Hilke:2010zz}, etc., has been crucial for progress in various fields like high energy physics, astronomy, and medical physics.
	Even though the geometry of an RPC is simple in comparison to other gaseous detectors, the physics processes associated with it (e.g., primary ionization, charge transport, electron multiplication, and signal formation) are as complex as any other.
	They are also conisdered reliable enough to be regularly used as timing, triggering, and tracking devices in numerous experiments \cite{Goswami_2017,Kumari_2020,Collaboration_2012,MONDAL2021166042}.
	A precise simulation model is needed to model RPCs and explain the experimental results adequately.
	A one-dimensional hydrodynamic model is beneficial for quick results at small computational time ~\cite{MOSHAII2012S168,Datta_2020,PFonte_2013}.
	This model works well when the number of space charges is significant, as the collective behavior of space charges meets the criteria of considering the system as a continuum \cite{Rout_2021}.
	A detailed Monte Carlo simulation model, where particles are traced individually, is the other option ~\cite{LippmanThesis, Lippmann_1, Lippmann:2003ar}.
	However, the large number of charges in the avalanches slows down the particle model significantly; hence it becomes computationally expensive. 
	\par Garfield++ \cite{Garfield,Veenhof:1993hz,VEENHOF1998726} is a C++-based simulation tool to simulate ionization detectors.
	It has its own geometry tools, which can be used to make simple virtual detectors.
	However, for complex geometries, one can borrow field maps and geometry files from external field solvers  like COMSOL \cite{comsol}, neBEM \cite{MAJUMDAR2008346,MAJUMDAR2009719} etc.
	Several methods to generate avalanches inside gaseous detectors in Garfield++, for example, microscopic tracking, Monte Carlo tracking, etc., are based on particle models.
	These tracking methods are very detailed but slow in a computational sense because all the algorithms run serially.
	One more drawback of tracking methods in Garfield++ is the absence of the dynamic space-charge effect, which is crucial when the number of space charges becomes significantly large ($\geq10^6$) so as to modify the applied field.
	\par To address the above issues of Garfield++, we have written an algorithm and introduced an extra class, pAvalancheMC, to generate avalanches with multithreading techniques using OpenMP \cite{OpenMp}.
	In this class, we have incorporated the space charge effect using the line charge model discussed in \cite{Dey_2020,Dey_2022}.
    Earlier a 2-D model was developed in \cite{LippmanThesis},  in which cylindrical co-ordinates $r$, $z$ and $\phi$ were used under the assumption of rotational symmetry
    	of the avalanche around the $z$-axis. Consequently, the present model can be
    	considered as a 3D Cartesian extension of the model developed in \cite{LippmanThesis, Lippmann_1, Lippmann:2003ar}. In our model we have not considered rotational symmetry.

	Another approach to parallelize Garfield++ using MPI can be found in \cite{pGarfield1,BOUHALI201892}, named pGARFIELD.
	However, pGARFIELD does not include the space-charge effect. 
	\par In Section \ref{sec:2_randomNumber}, we describe the method of generation of uniform and uncorrelated parallel random numbers using the TRandom3 class of ROOT \cite{root-cern}, which is the heart of the Monte Carlo simulations.
	Similarly, use of OpenMP and FastVol to accelerate electric field computation are described in Section \ref{sec:3_neBEM}.
	Section \ref{sec:4_stepsOfAvalanche} describes the steps or algorithm of generation of a Monte Carlo avalanche, including the dynamic space-charge effect.
	In Section \ref{sec:5_instanceAvalancheConst}, we provide an example of avalanche saturation with a gas mixture of 80\% Ar and 20\% $\ce{CO_2}$ in an 2 mm gas gap RPC, generated through pAvalancheMC.
	In Section \ref{sec:examplewithsf6}  we provide another example of avalanche saturation and streamer formation with a gas mixture of 97\% $\ce{C_2H_2F_4}$, 2.5\% $\ce{i-C_4H_{10}}$, and 0.5\% $\ce{SF_6}$. in an 2 mm gas-gap RPC, generated through pAvalancheMC. 
  It may be
  mentioned that the results obtained in this work are quite similar to those in \cite{LippmanThesis, Lippmann_1, Lippmann:2003ar}. In all these computations similar physics processes such as radial
  blow-up and contraction of the avalanches have been observed. Despite the
  differences between the two models, \textcolor{black}{it is encouraging to note that} the order
  of magnitude of space-charge effect is comparable. In Section \ref{sec:6_speedUp}, the performance of the multithreaded code has been discussed.
	Section \ref{sec:7_neBEM} compares induced-charge \textcolor{black}{distributions} for three different voltages 1720 V, 1730 V and 1735 V with space-charge effect, calculated using neBEM and the proposed version of Garfield++.

	\section{Uncorrelated and thread safe parallel Random number generation using Trandom3 and OpenMP} \label{sec:2_randomNumber}
		
	We know that the outcome of any Monte Carlo simulation depends on the quality of the generated random numbers.
	It is important that the generated random numbers are uncorrelated, and the corresponding random engine has a considerable period.
	TRandom3 is a class of ROOT \cite{BrunRoot}, that generates uniform random numbers.
	It is based on the Mersenne Twister (MT) algorithm and provides a period of $2^{19937}-1$  and a 623-diminsionally equidistributed uniform pseudorandom number generator up to 32-bit accuracy \cite{MersenneTwister}.

\begin{figure}
		\center{\includegraphics[width=0.6\linewidth]{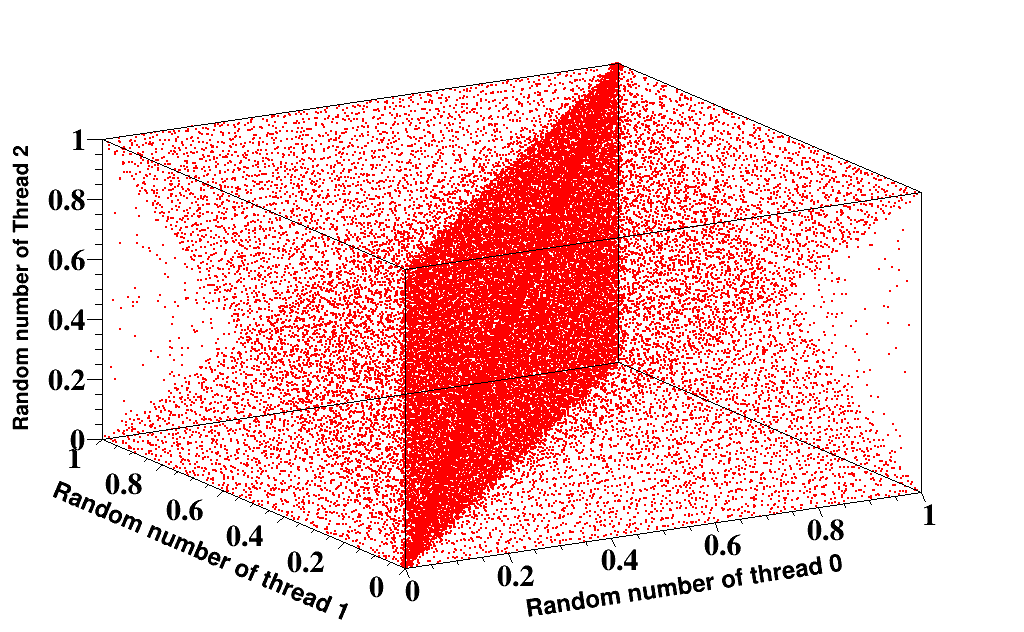}}
		\caption{Correlation between random numbers generated in parallel using the same seed by threads 0,1,2.}
		\label{fig:Correlated}
	\end{figure}

	We have chosen TRandom3 as the random engine in our simulation code.
	The random numbers generated 
	serially using TRandom3 are uncorrelated. 
	However, we have seen that the former engine leaves some correlation between random numbers generated in parallel in different threads (see Fig. \ref{fig:Correlated}) when all threads use the same random engine and the same seed.
	
	The random numbers corresponding to different threads, namely 0, 1, and 2, generated from the same object of the random engine are strongly correlated and mostly lying on three planes (see Fig. \ref{fig:Correlated}).

	\par
\begin{figure}
		\center{\includegraphics[width=0.6\linewidth]{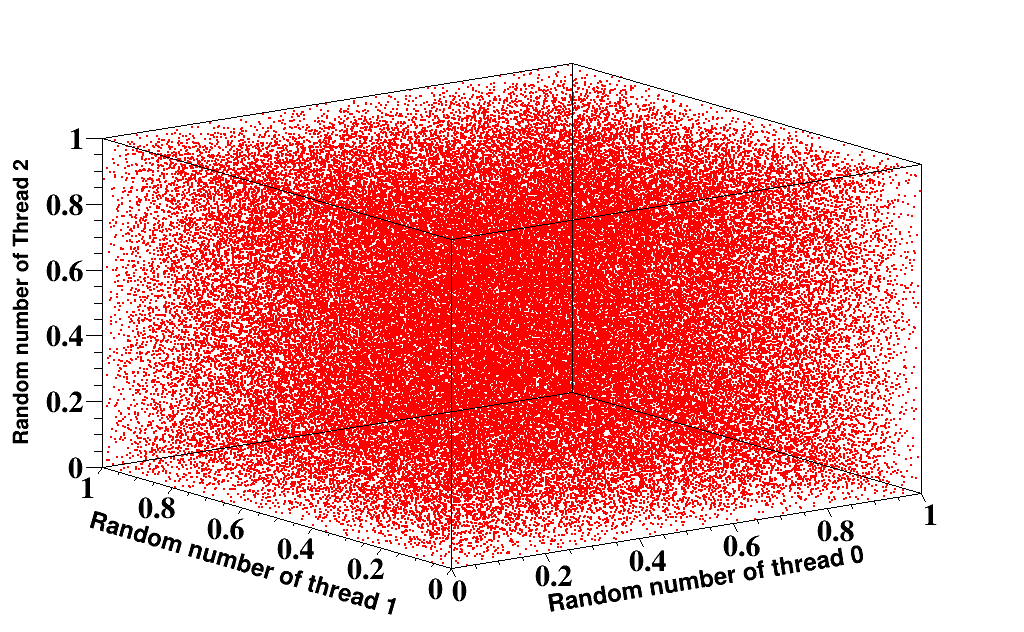}}
		\caption{Correlation between random numbers generated in parallel using different seeds by threads 0,1,2.}
		\label{fig:unCorrelated}
	\end{figure}

 To solve this problem,  objects of the TRandom3 class have been assigned for three threads with three different seed values.
The 3D correlation plot using random numbers corresponding to threads 0, 1 and 2 (see Fig.~\ref{fig:unCorrelated}) is uniformly distributed over the volume. 

	\par The method outlined above is implemented in class pAvalancheMC. Using the function SetNumberOfThreads(n), one can specify the number of threads used for random number generation.
	The different objects of the TRandom3 class for different threads can be assigned using the same function.
	Using a unique thread ID created by OpenMP, all threads can generate random numbers from their corresponding objects.
	An example of C++ code for the usage of the function SetNumberOfThreads(n) is given below:
	\vspace{2.5cm}	
	\begin{lstlisting}%[frame=none]
		vector<TRandom3 *>rnd;
		void SetNumberOfThreads(int nThread)
		{
			rnd.resize(nThread);
			for(int ThreadId=0; ThreadId<nThread; ThreadId++)
			rnd[ThreadId]= new TRandom3(0);
		}
	\end{lstlisting}
	
	\section{Accelerated electric field solution using OpenMP and FastVol}
	\label{sec:3_neBEM}
	
	The Open Multi-Processing (OpenMP) is an Application Programming Interface (API) that supports multi-platform shared memory multiprocessor programming in C, C++ and FORTRAN on most processor architectures and operating systems \cite{OpenMp}.
	It consists of a set of compiler directives, library routines and environment variables that influence runtime behavior and uses a simple, scalable API for developing parallel applications on platforms ranging from the standard desktop to supercomputers.
	We have successfully implemented OpenMP for the neBEM field solver \cite{MAJUMDAR2009719}.
	The parallelization has been implemented in several sub-functions of the toolkit, such as computation of the influence coefficient matrix, matrix inversion and evaluation of the field and potential at desired locations.
	These routines are computation intensive since there can be thousands of elements where the charge densities need to be evaluated and the influence of all these elements need to be taken
	into account. 
	The task is further complicated due to the use of repetitive structures, in order to conform to the real geometry of a detector.
	This has proved to be very important in improving the computational efficiency of the solver.
	We have tested these implementations on up to 24 cores.
	The observed reduction in the computational time is significant while the precision of the solution is preserved.
	\par Even after adopting OpenMP for solving for the charge distribution on all the material interfaces of a given device, the time needed to calculate potentials and fields for a complex device can become prohibitive.
	This is especially so if the device (not usually true for an RPC) is composed of hundreds of primitives, thousands of elements and several tens of repetitions.
	Reduction of time taken to compute electrostatic properties becomes increasingly important when complex processes such as avalanche, Monte-Carlo tracking and micro-tracking are being modelled using Garfield++.

	In order to model these phenomena within a reasonable time, we have implemented the concept of using pre-computed values of potential and field at large number of nodal points in a set of suitable volumes.
	These rectangular volumes are chosen such that they can be repeated to represent any region of a given device and 	simple trilinear interpolation is used to find the properties at non-nodal points.
	The associated volume is named as the Fast Volume (FastVol) which may be staggered, if necessary.
	In order to preserve the accuracy despite the use of trilinear interpolation, it is natural that the nodes should be chosen such that they are sparse in regions where the potential and fields are changing slowly and closely packed where these properties are changing fast.
	
	\section{Steps of an avalanche simulation with space-charge effect in Garfield++}
	\label{sec:4_stepsOfAvalanche}
	
	This section will discuss techniques for implementing the space-charge effect with multithread in the pAvalancheMC class of Garfield++.
	In further discussions, we will use the short name $E_a$ as the applied field, $E_s$ as the space-charge field, and $E_t$ as the sum of applied and space-charge fields.
	
	The necessary steps to simulate an avalanche are given below:
	\begin{enumerate}[(i)]
		\item
		Selection of primaries to start avalanche according to their generation time.
		\item
		Parallel calculations of drift line of avalanche charges.
		\item
		\textcolor{black}{Calculation of gain in each step}.
		\item
		Generation of cylindrical grid of space charge region.
		\item
		Calculation of dynamic space-charge field.
	\end{enumerate}
	
	\subsection{\textbf{Selection of primaries to start avalanche according to their generation time}}
	When a charged particle passes through a gas detector, it can interact with gas molecules and leave some primary ionizing particles along its path.
	A track of a charged particle through the gas gap and corresponding primary ionizations can be simulated using the class TrackHeed \cite{SMIRNOV2005474} of Garfield++.
	The locations and time of generation of primary clusters are then stored in an array.
	The primaries (here electrons) can be generated at different instants of time along the track.
	Hence, the starting time of the avalanche generation from different primaries can be different.
	Therefore, if we arrange primaries in ascending order based on the generation time, then depending on the simulation step $\delta t$, they will start the avalanche process.
	For example, if $t$ is the generation of time of any primary electron, it will start an avalanche when $n\delta t\geq t$, where $n$ is the number of steps completed in the simulation. 
	
	\subsection{\textbf{Parallel calculations of drift line of avalanche charges}}
	The AvalancheMC class of Garfield++ can simulate the drift path of charges serially.
	Hence, it becomes slow when the number of charges is large.
	To solve this issue, we assign the secondary particles to the available threads using OpenMP and propagate them in the gaseous medium simultaneously at every step of time to calculate further drift points.
	The diffusion part of pAvalancheMC is kept the same as in-class AvalancheMC, where the thermal diffusion is considered Gaussian and anisotropic when an electric field is present~\cite{LippmanThesis}.
	Therefore, the diffusion will be in two directions, longitudinal and transverse.
	For total electric field $E_t$, the drift velocity ($v_D(E_t)$), longitudinal ($D_L(E_t)$) and transverse diffusion ($D_T(E_t)$) constants have been calculated by using MAGBOLTZ \cite{BIAGI1989716,BIAGI1999234}.
	The variance of longitudinal Gaussian distribution is
	$\sigma_L=D_L\sqrt{v_D\delta t}$ and that of transverse diffusion is $\sigma_T=D_T\sqrt{v_D\delta t}$.   
	
	\subsection{\textbf{Calculation of Gain in each step}}
	{\color{black}The calculation of gain in the pAvalancheMC class is the same as in the AvalancheMC class. It is based on the Yule-Furry model and includes the effect of electron attachment \cite{Furry,Schindler:1500583}}.
	At each step, the probability of ionisation and attachment is calculated using the Townsend ionisation ($\alpha$) and attachment ($\eta$) coefficients, where $\alpha$ and $\eta$ are computed using MAGBOLTZ \cite{BIAGI1989716,BIAGI1999234}.

	\subsection{\textbf{Calculation of dynamic space-charge field}}
		 
	The space-charge effect is turned on when the number of electrons crosses a threshold.
	This threshold is determined by the user and, for the computations presented in this paper, the value has been set to $10^4$.
	The generation of grid elements and finding locations of the space charges in the grids have been divided into four steps.
	The steps are as follows:\\
		
\begin{figure}[H]
		\center\subfloat[\label{fig:zGrid}]{\includegraphics[scale=0.4]{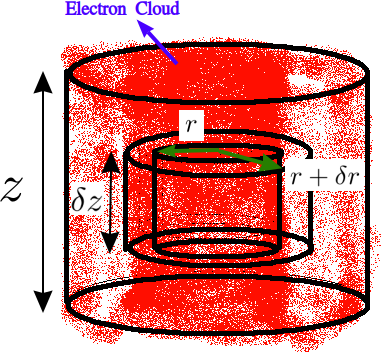}}\subfloat[\label{fig:rGrid}]{\includegraphics[scale=0.5]{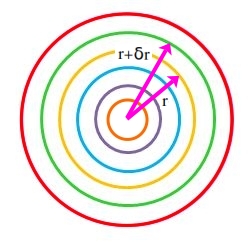}}
		
		\center\subfloat[\label{fig:cGrid}]{\includegraphics[scale=0.1]{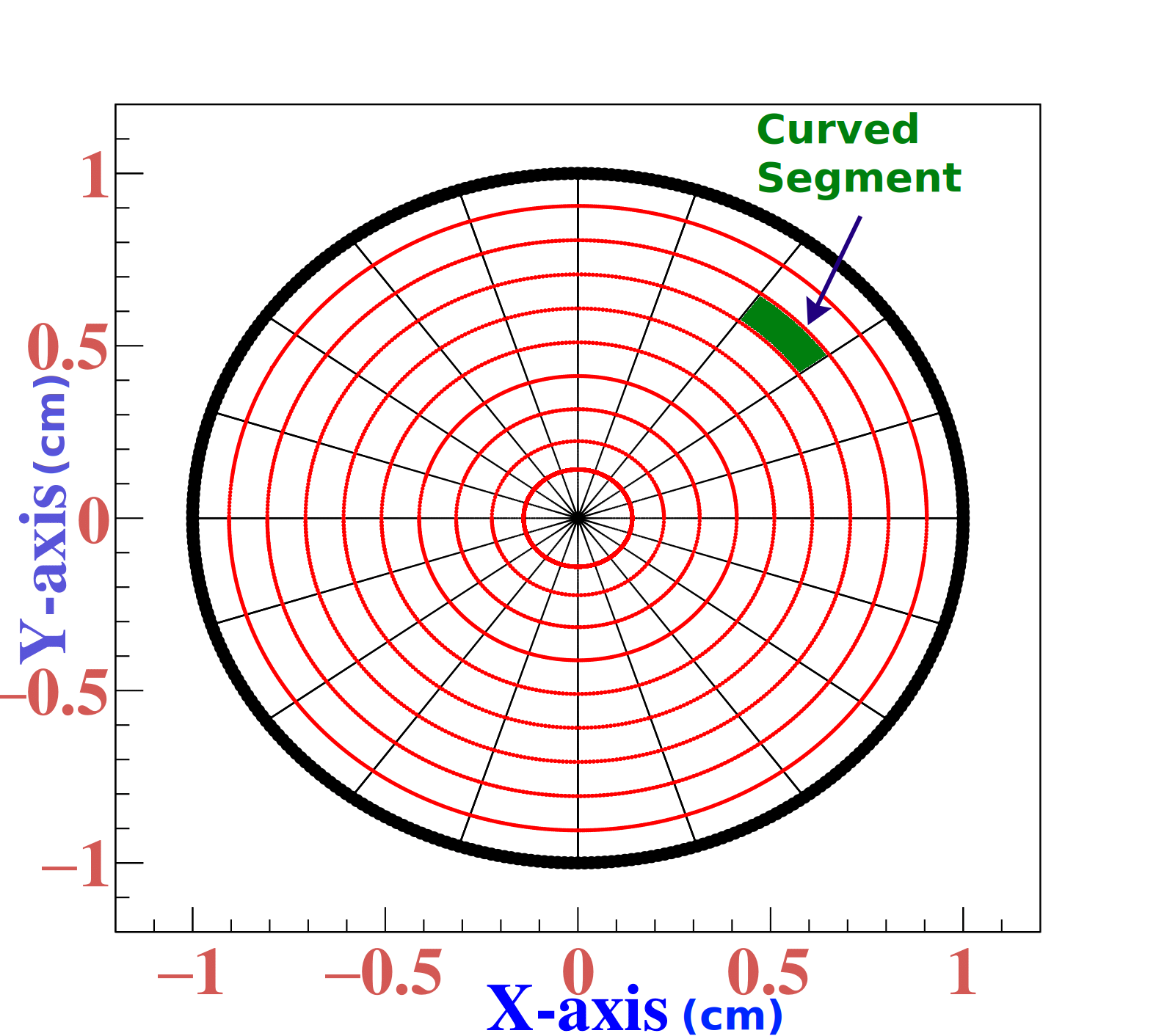}}\subfloat[\label{fig:sGrid}]{\includegraphics[scale=0.35]{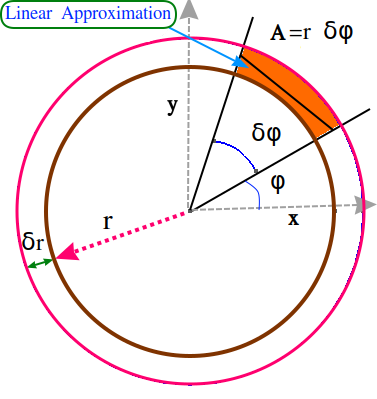}}
		\caption{Steps to generate cylindrical grid, (a) dividing z space into elements of size $\delta z$, (b) radial space divided into several co-centric rings of increasing radius r and thickness $\delta r$, (c) divide rings into several small curved segments, (d) decrease size of curved elements and approximation by straight line.}
	\end{figure}

	Step  1: Let us consider at a particular time step of a growing avalanche, the space charges are distributed as shown in Fig. \ref{fig:zGrid}.
	To calculate the field for this charge distribution, the $z$-directional space is divided into $S$ elements of size $\delta z$, as shown in Fig. \ref{fig:zGrid}. 
		
	Step  2: For a particular $z$ position, \textcolor{black}{the avalanche region is divided into R co-centric rings each having} a thickness $\delta r$ as shown in Fig. \ref{fig:rGrid}.
	For all $z$, the thickness $\delta z$ is considered as small as 0.001 cm.
	
	Step  3: In the third step, each ring can be divided into $L$ curved segments (see Fig. \ref{fig:cGrid}).
	Hence, each curved segment will lie between angles of $\phi$ and $\phi+\delta \phi$, where $\phi$ is the azimuthal angle subtended to the center of the circle or ring (see Fig. \ref{fig:sGrid}).
	Therefore, the $\phi$ directional space will also be divided into $L$ segments of size $\delta \phi$, so that $L\delta \phi=360$\textdegree \space(for more details see ~\cite{Dey_2020}).
	
	Step  4: In this calculation, $\delta \phi$ is considered very small ($\approx 1$\textdegree) so that each curved segment can be considered as a straight line of length $r\delta \phi$ (see Fig. \ref{fig:sGrid}).
	
    Therefore, the entire avalanche volume is segmented into small volumes (voxel). The charges within a specific voxel is represented by a line charge with constant charge density situated within that volume.
     The total number of voxels  $T=S\times R\times L$.
     
	\par The space-charge field at any voxel is the sum of the electric fields due to other voxels containing non-zero charges.
	The space-charge field is considered to be uniform within a voxel.
	The total field at any location ($r,\phi,z$) is the sum of the applied field and space-charge field.
	Since voxels are considered as line segments, the line charge formula has been used to calculate the space-charge field, as discussed in ~\cite{Dey_2020,Dey_2022}. 
	
	\subsection{\textcolor{black}{\textbf{Implementation of positive and negative ions}}}
	\textcolor{black}{Positive ions are considered to be generated at the ionization locations. The attachment positions of the electrons within the gas gap are considered as the positions of the singly charged negative ions. It may be noted that, since the electron drift velocity is much higher than that of the ions, the ions are approximated to be static in this paper. However, implementation of ion drift is possible, if necessary,  by providing additional information about ion mobility.}
	
	\section{Instance of a simulated avalanche inside an RPC with space-charge effect in Ar and $\ce{CO_2}$ gas-mixture  }\label{sec:5_instanceAvalancheConst}
	\subsection{\textcolor{black}{\textbf{Avalanche in absence of the negative ion effect}}} \label{subsec:aval_cor_nonegions}
	This section will discuss an example of an avalanche simulated in an RPC using pAvalancheMC with and without the space-charge effect.
	A gas mixture of 80\% Ar and 20\% $\ce{CO_2}$ is used.
	The electrodes of the RPC are made of a 2 mm thick Bakelite with an area of $30\times 30~\text{cm}^2$.
	The gas gap is also taken as 2 mm.
	A constant electric field along $+z$ ($E_{a|z}$) of 23.5 kV/cm has been applied across the gas gap with the help of ComponentConstant class of Garfield++.
	The time-step of the simulation is taken as 20 ps.
	\textcolor{black}{Muon tracks and primary ionisation inside the gas gap are generated} using HEED \cite{SMIRNOV2005474}. \textcolor{black}{It is noted that the contribution of the negative ions are not considered while performing this avalanche}. 
	\par
\begin{figure}[H]
		\center\subfloat[\label{fig:gain}]{\includegraphics[width=0.5\linewidth]{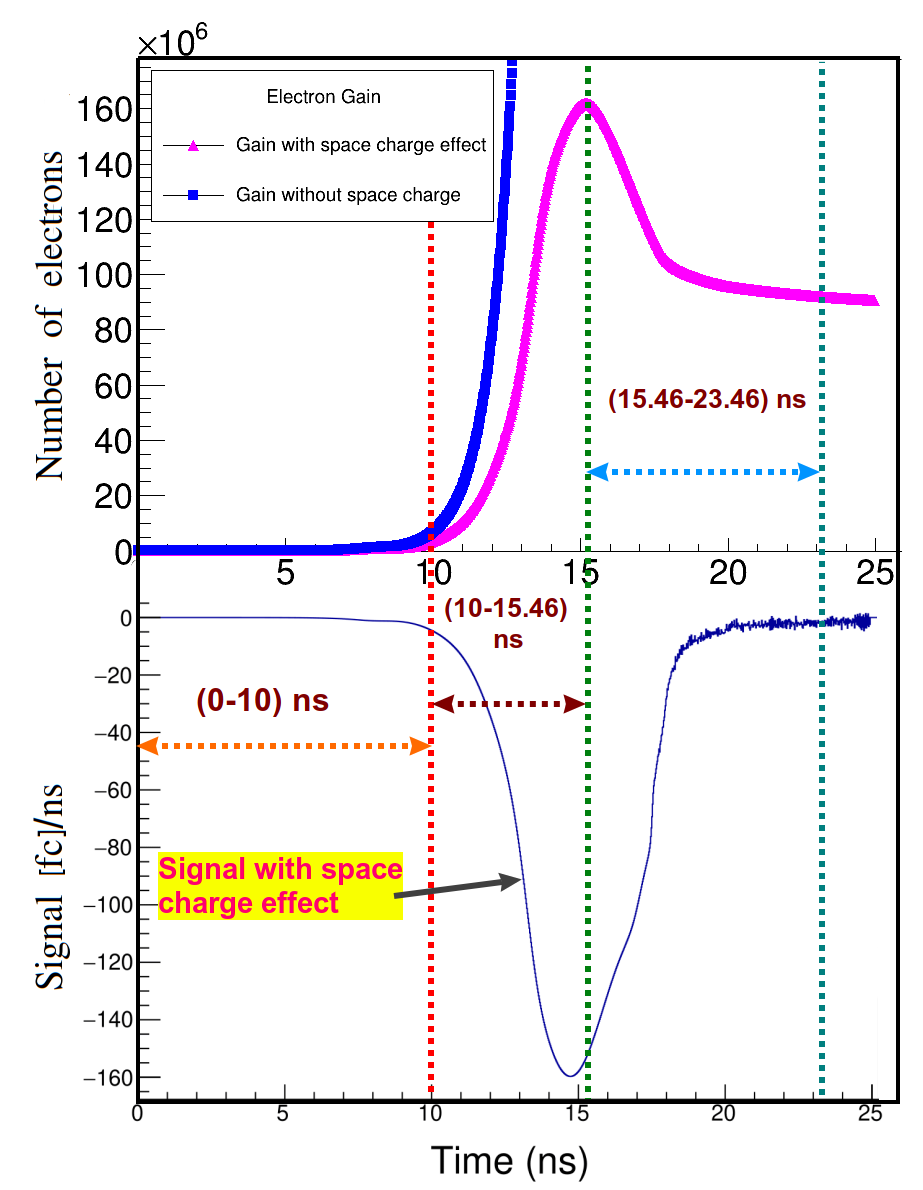}}\subfloat[\label{fig:max_eField_arco2}]{\includegraphics[width=0.6\linewidth]{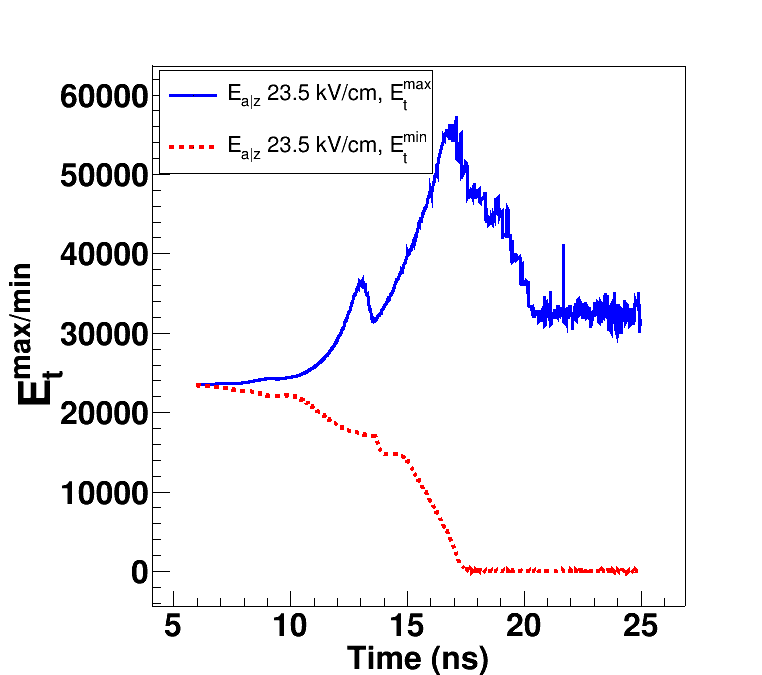}}
		\caption{(a) Variation of number of electrons and signal with time, \textcolor{black}{(b) variation of total electric field $E_{t}^{max}$ and $E_{t}^{min}$ with time in absence of negative ions.}}
	\end{figure}

	In  Fig. \ref{fig:gain}, the electron gain at each step of time with and without the space-charge effect is compared.
	The blue and pink curves specify the electron gain without and with the space-charge effect, respectively.
	For avalanche times below 10 ns, the two curves almost overlap.
	After 10 ns, the avalanche charge without space-charge effect continuously grows, while the avalanche charge with space-charge effect (pink curve) grows slowly, reaches a peak at around 15.46 ns, and then starts showing a saturation effect after 17.96 ns\textcolor{black}{, at an electron count of approximately $10^8$}.
	
		\begin{figure}[H]
		\center\subfloat[\label{fig:gain_negeffc}]{\includegraphics[width=.47\linewidth]{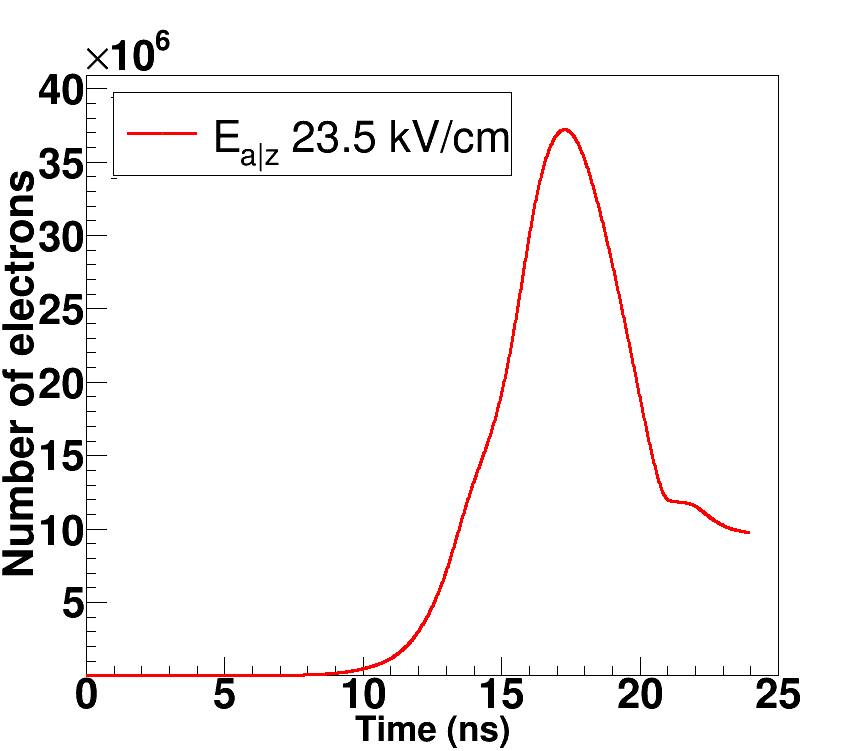}} \subfloat[\label{fig:et_negeffc}]{\includegraphics[width=.54\linewidth]{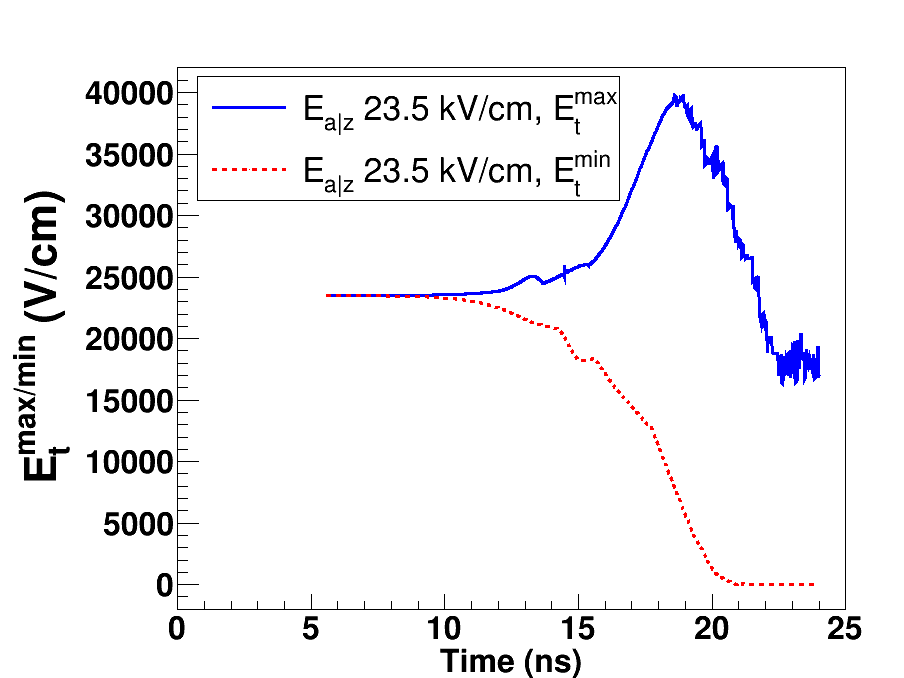}}
		
		\subfloat[\label{fig:ratio_neg_ion}]{\includegraphics[width=.52\linewidth]{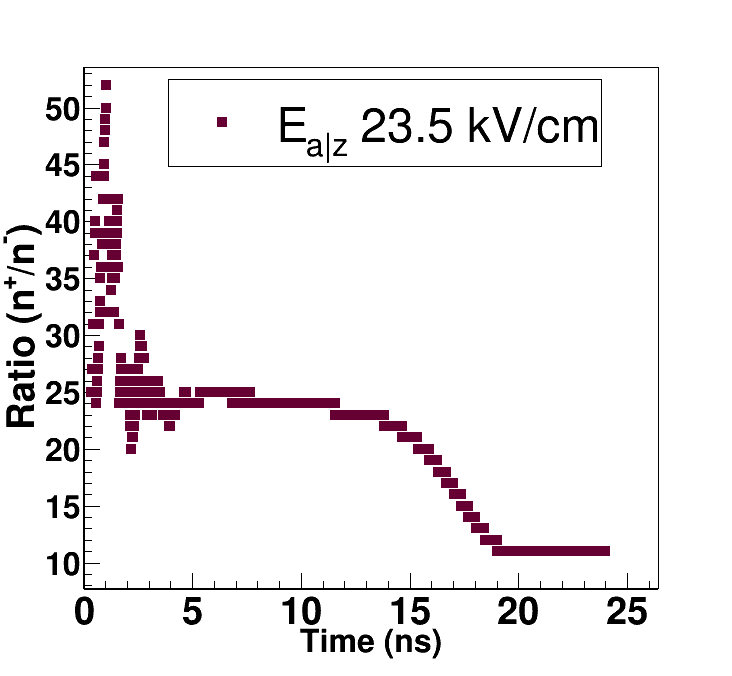}}\subfloat[\label{fig:alpha_eta}]{\includegraphics[width=.50\linewidth]{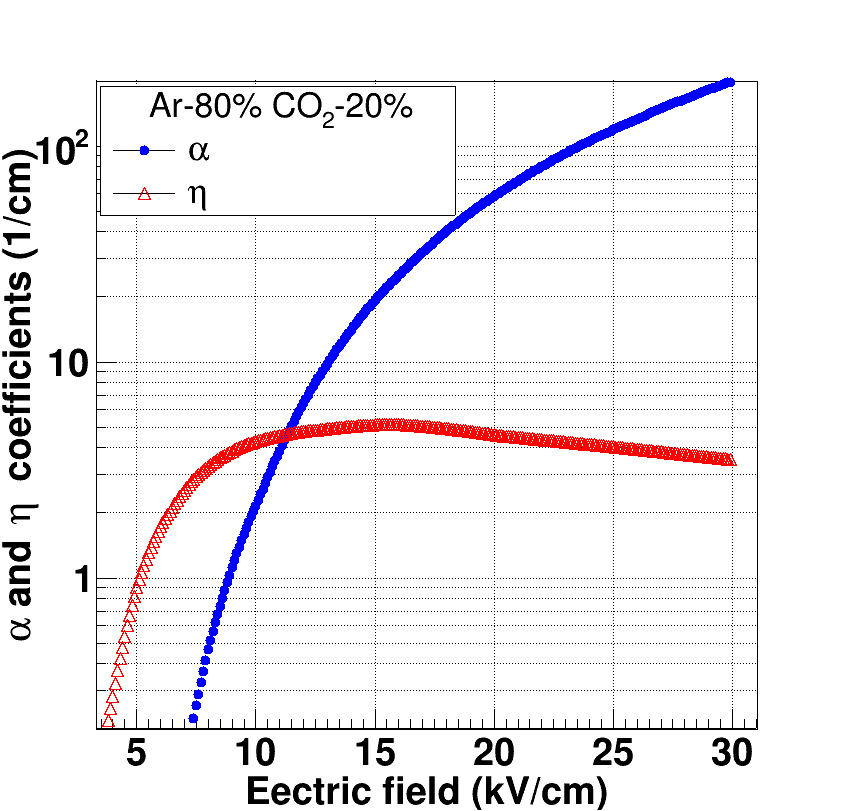}}
		\caption{\textcolor{black}{(a) Variation of number of electrons with time in presence of negative ions, (b) Variation of total electric field $E_{t}^{max}$ and $E_{t}^{min}$ with time in presence of negative ions, (c) variation of ratio between number of positive ions ($n^{+}$) and  negative ions ($n^{-}$) with time for a gas mixture 80\% of Ar and 20\% of \ce{CO}$_2$, (d) variation of ionisation ($\alpha$) and attachment ($\eta$) coefficients with electric field for a gas mixture  80\% of Ar and 20\% of \ce{CO}$_2$.}}
	\end{figure}
		\subsection{\textcolor{black}{\textbf{Avalanche in presence of the negative ion effect}}} \label{subsec:aval_negion_arco2}
	\textcolor{black}{ The configurations of the simulation are the same as those in subsection \ref{subsec:aval_cor_nonegions}. The variation in the number of electrons with time in the presence of the negative ions is shown in Fig. \ref{fig:gain_negeffc}, and the variation in the maximum of the total electric field $E_t^{max}$ is shown in Fig. \ref{fig:et_negeffc}. A result from the comparison of the time variation in the number of electrons with the space charge effect of Figs. \ref{fig:gain} (pink curve) and \ref{fig:gain_negeffc} indicates that the gain has been reduced by approximately a factor of four.} \par\textcolor{black}{As in Fig. \ref{fig:gain}, saturation in the variation of the number of electrons with time can be observed in Fig. \ref{fig:gain_negeffc}. However, the knee point or starting point of the saturation region for the pink curve in Figs. \ref{fig:gain} and \ref{fig:gain_negeffc} is at approximately 17.6 ns and 21 ns, respectively. The number of electrons at the knee point of Fig. \ref{fig:gain} is approximately 8.7 times the number of electrons at the knee point of Fig. \ref{fig:gain_negeffc}. Moreover, it is found that the saturation curve of Fig. \ref{fig:gain_negeffc} is steeper than the pink curve of Fig. \ref{fig:gain}. This is because, in the presence of negative ions, the maximum electric field in the saturation region lies between 17-20 kV/cm and minimum electric field is close to zero, as shown in Fig. \ref{fig:et_negeffc}, whereas in the absence of negative ions, it was beyond 30 kV/cm as shown in Fig. \ref{fig:max_eField_arco2}. Hence, the gain factor is reduced. } \par
	\textcolor{black}{The variation in the ratio of the number of positive ions ($n^{+}$) to negative ions ($n^{-}$) is illustrated in Fig. \ref{fig:ratio_neg_ion}. From the previous figure, it becomes evident that prior to 10 ns, the ratio fluctuates between 20 and 52. After 10 ns, the ratio gradually decreases, and it stabilizes at approximately 11 after 18 ns. This substantial imbalance between negative and positive ions can be attributed to the difference of magnitude between attachment ($\eta$) and ionization ($\alpha$) coefficients, as depicted in Fig. \ref{fig:alpha_eta}. It is apparent from Fig. \ref{fig:alpha_eta} that at around 12 kV/cm, $\alpha$ and $\eta$ attain equal values. As the electric field increases, the difference between $\alpha$ and $\eta$ increases rapidly. Therefore, below 12 kV/cm, attachment dominates over ionization. As previously mentioned, in the saturation region, the maximum electric field in certain parts of the avalanche region reaches around 20 kV/cm. At this point, the difference between $\alpha$ and $\eta$ is approximately 55 1/cm, indicating that the ionisation process prevails. Consequently, in the saturation region, the number of $n^{+}$ and $n^{-}$ becomes balanced, resulting in a constant ratio. }

	\section{\textcolor{black}{Visualisation of transport of electrons under space charge effect}} 
	\textcolor{black}{In order to understand the transport of charged particles through the gas volume, the distribution of $z$-positions vs. radial positions of the electrons at different instants of avalanche growth have been analyzed ( Fig. \ref{fig:Before10ns}). The following discussion contains the result of the analysis. }
	\par\textcolor{black}{ It is evident from subsection \ref{subsec:aval_negion_arco2} that the Ar-CO$_2$ gas mixture does not exhibit a strong attachment coefficient, resulting in a higher concentration of positive ions compared to negative ions. Therefore, for the sake of expeditious computation and simplicity, we neglect the effect of negative ions and focus solely on the transport of electrons.}
   \par \textcolor{black}{	 We can divide the pink curve of Fig. \ref{fig:gain} into three regions: a) before 10 ns, where space-charge effects are negligible, b) from 10 ns to 15.46 ns, when gain increases rapidly, and c) after 15.46 ns, where the gain drops and maintains a saturated value.}
	\subsubsection{ \textbf{From 0 ns to 10 ns, overlap region of gain of Fig. \ref{fig:gain}}}
\begin{figure}
		\center\subfloat[\label{fig:0ns}]{\includegraphics[scale=0.17]{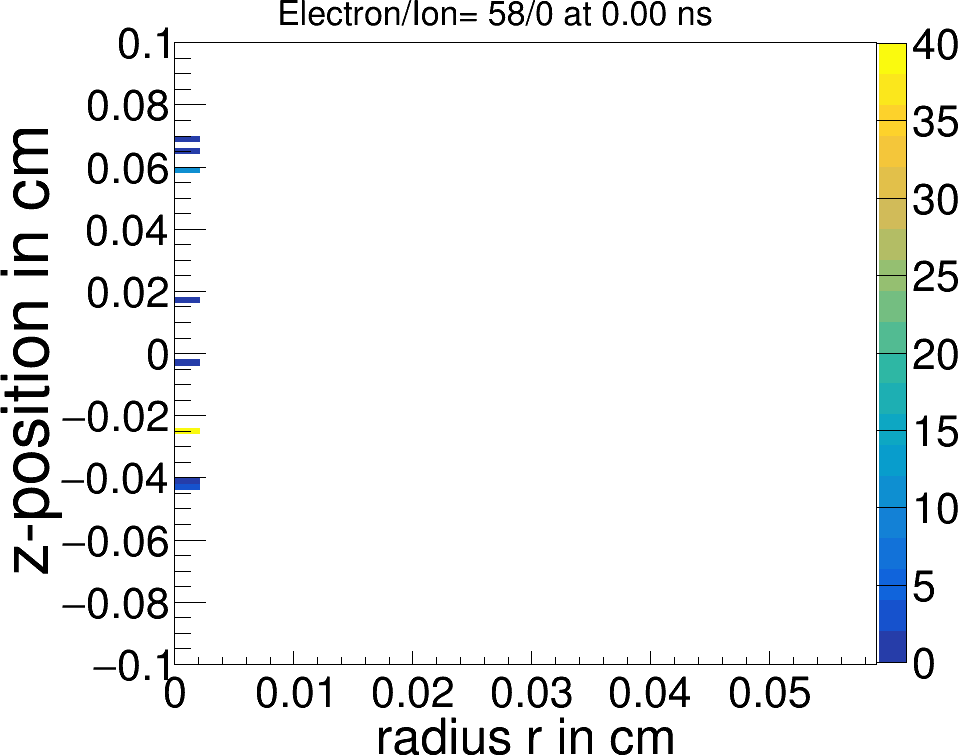}
			
		}\subfloat[\label{fig:4.5ns}]{\includegraphics[scale=0.17]{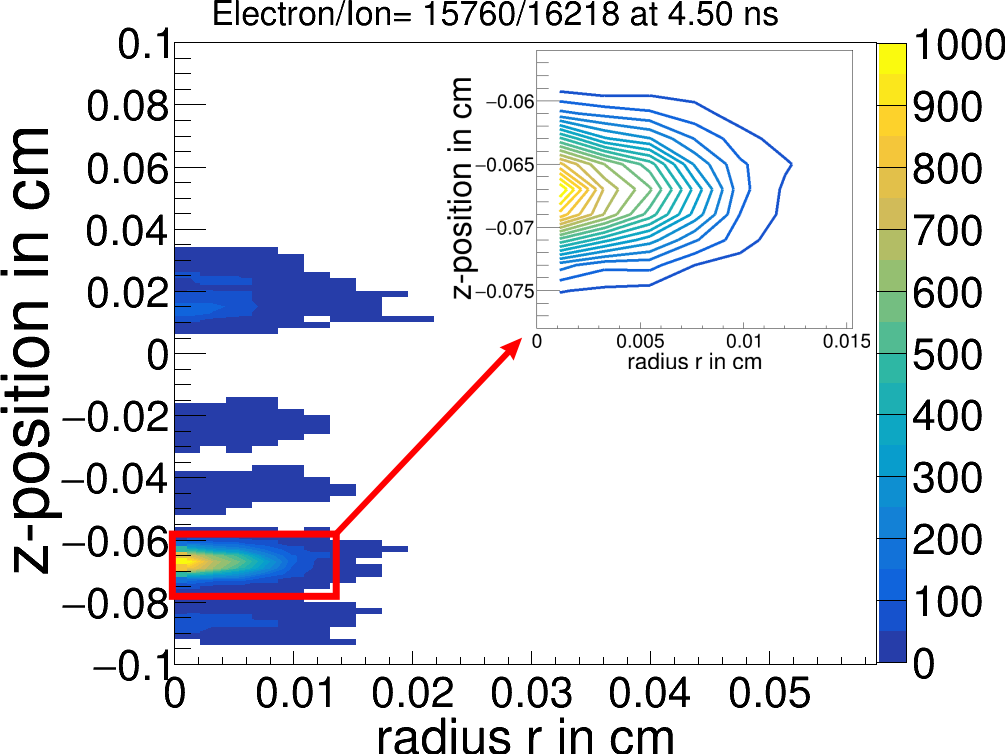}
			
		}
		
		\center\subfloat[\label{fig:7.46ns}]{\includegraphics[scale=0.17]{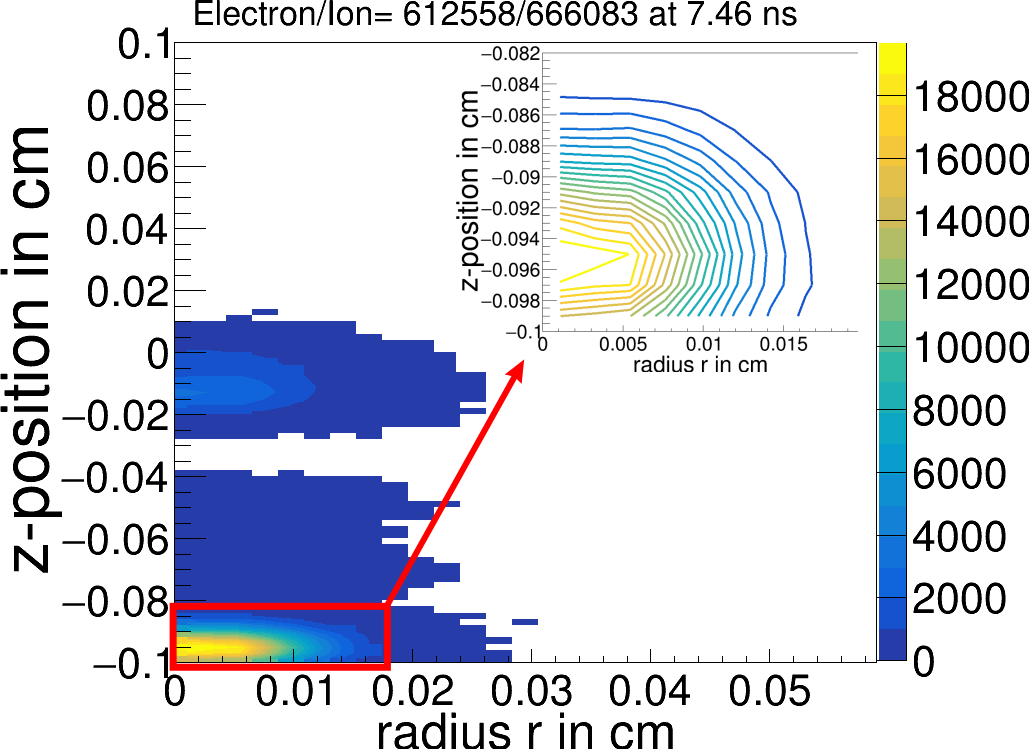}
			
		}\subfloat[\label{fig:10.46}]{\includegraphics[scale=0.17]{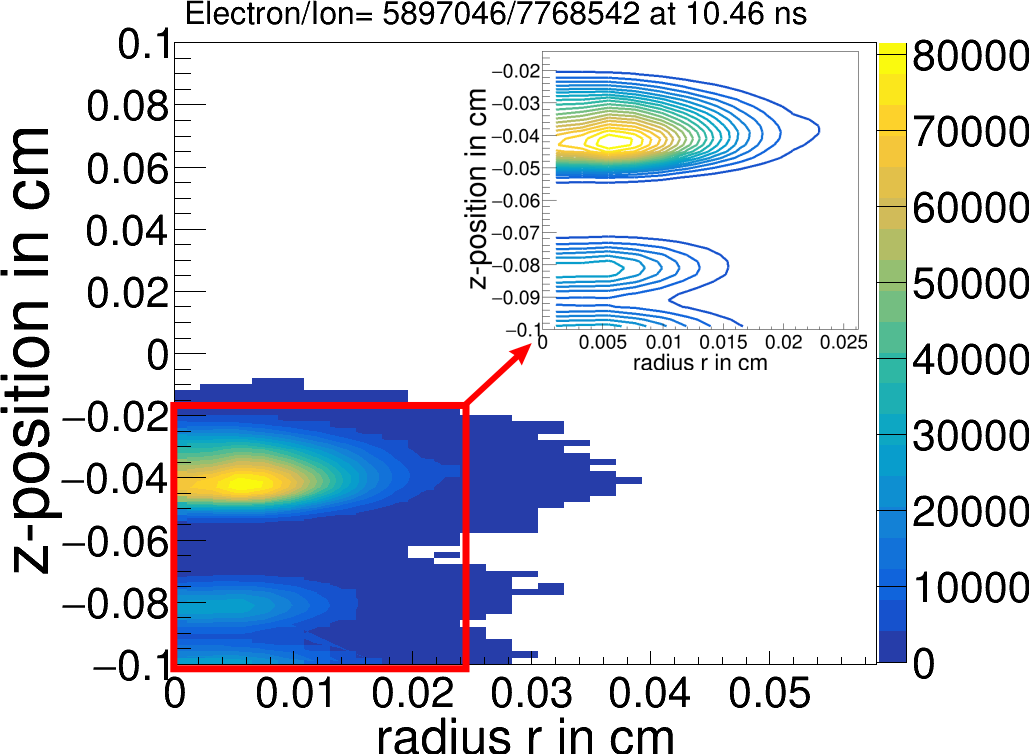}}
		
		\caption{Location of electrons in $z-r$ plane at time (a) 0. ns, (b) 4.5 ns, (c) 7.46 ns and (d) 10.46 ns. The inlet shows a zoom into the highlighted electron cluster.}\label{fig:trk_beforesp}
	\label{fig:Before10ns}	
	\end{figure}

	\begin{enumerate}[i.]
		\item
		At time 0 ns (see Fig. \ref{fig:0ns}) there are in total 58 electrons distributed in 7 clusters. 	A small number of primary ions are generated along with the primary electrons. They are neglected since they do not participate strongly in the rest of the evolution that is being followed in the present study.
		It is noted that $z$=0.1 is the anode plane and $z$=-0.1 is the cathode plane.
		The color bar represents the number of electrons in a ring of radius r and thickness $\delta r$.
		\item
		In Fig. \ref{fig:4.5ns}, it is shown that at time 4.5 ns electrons have already started participating in the ionization process, and a total of 15,760 electrons and 16,218 ions are generated.
	    In this figure, two clusters near the cathode plane at $z$=-0.1 merge and form a big cluster.
		Therefore, the total number of electron clusters at this stage is 4.
		The maximum charge density is in the middle of the distribution (at radius $r=0$).
		The electron distribution is roughly symmetric. 
		\item
		In Fig. \ref{fig:7.46ns} at time 7.46 ns, there are only two electron clusters. The total number of electrons and ions is 612,558 and 666,083, respectively.
		Also, the charged center of the merged distribution has reached the anode, and the electron density contour is smooth and symmetric.
	\end{enumerate}	

	\begin{figure}
	\center\subfloat[\label{fig:10.46ns_z}]{\includegraphics[scale=0.2]{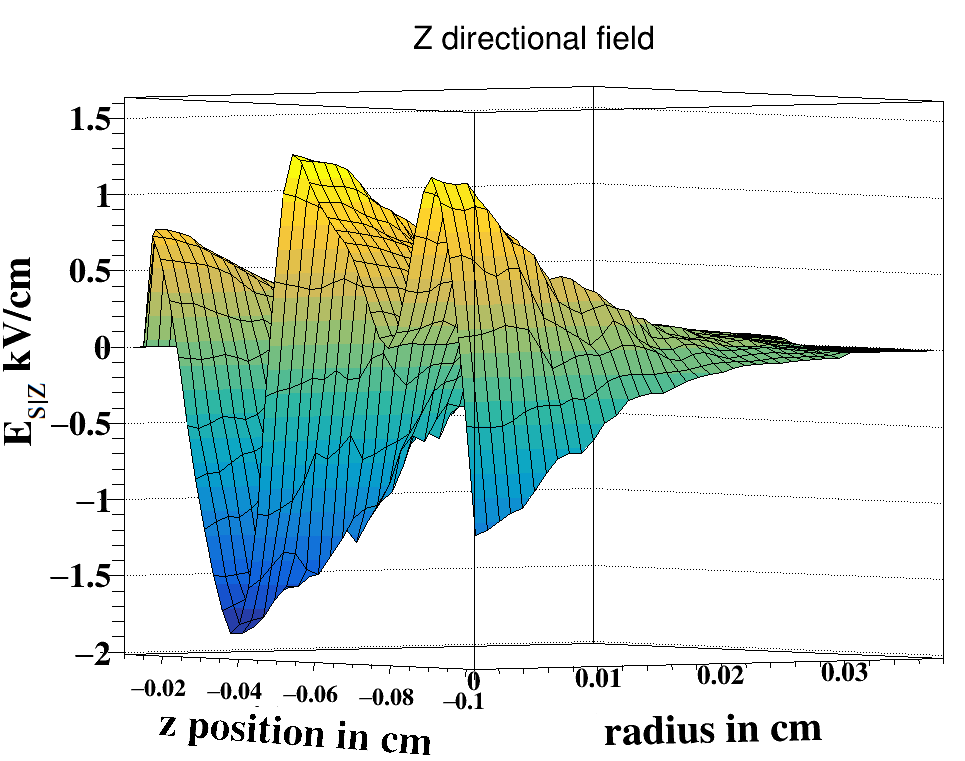}}\subfloat[\label{fig:10.46ns_rad}]{\includegraphics[scale=0.2]{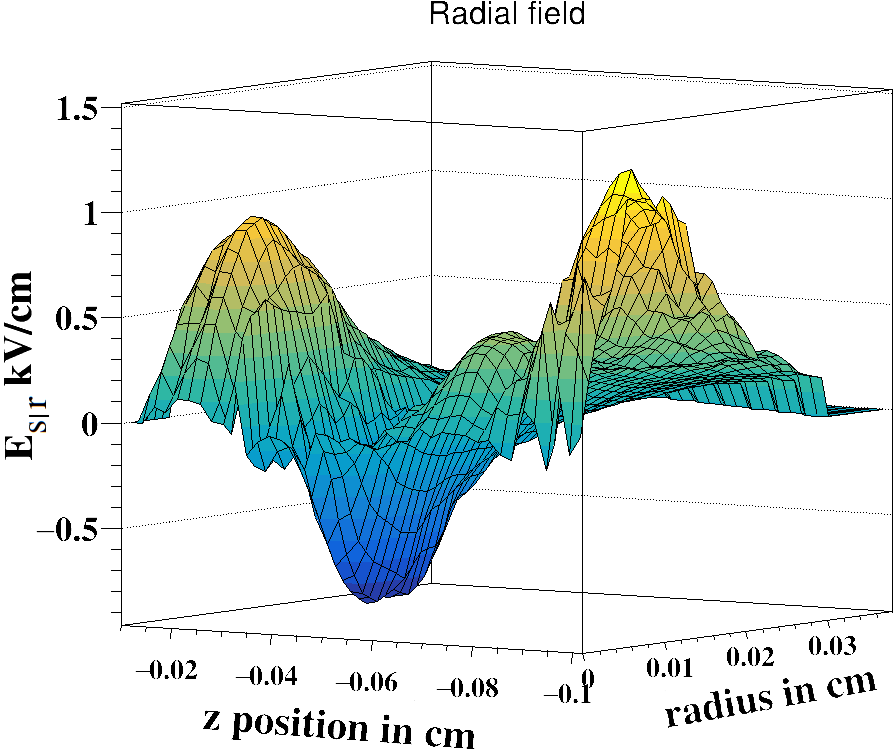}}
	
	\center\subfloat[\label{fig:10.46ns_phi}]{\includegraphics[scale=0.25]{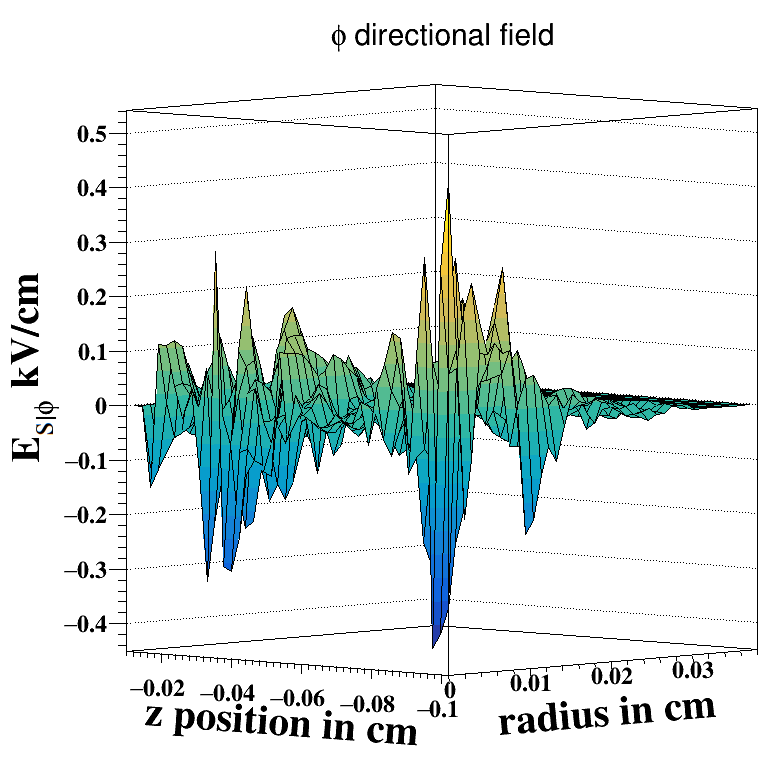}	
		
	}~\subfloat[\label{fig:10.46nsE}]{\includegraphics[scale=0.27]{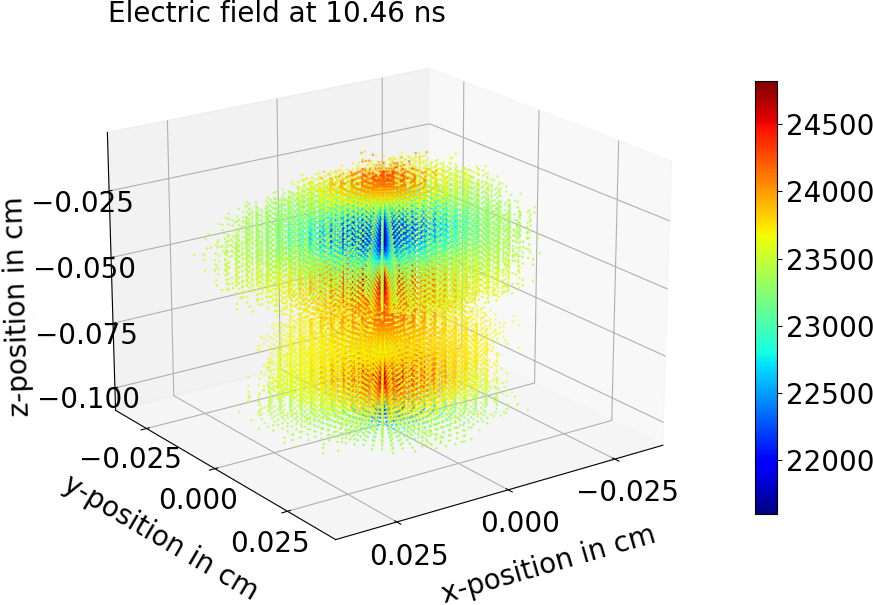}}
	
	\caption{(a) $z$-component of space-charge field ($E_{s|z}$) at 10.46 ns, (b) Radial component of space-charge field ($E_{s|r}$) at 10.46 ns, (c) $\phi$-component of space-charge field ($E_{s|\phi}$) at 10.46 ns, (d) Shape of the electron cloud and total electric field ( $E_{t}$) at different voxels at time 10.46 ns.}
\end{figure}
	
	\subsubsection{ \textbf{From 10 ns to 15.46 ns, till peak region of Fig. \ref{fig:gain}}}
	\begin{enumerate}[i.]
		\item
		At 10.46 ns, the blue and pink curves of Fig. \ref{fig:gain} start to separate, which signifies the relative importance of the space-charge effect.
		There are 5,897,046 electrons and 7,768,542 ions (see Fig. \ref{fig:10.46}).
		At this stage, the last two clusters also merge and form a single big cluster.
		A very close look at Fig. \ref{fig:10.46} shows that the charge center of the last cluster from the anode has not remained at $r=0$; rather, it shifted to the right.
		This is because $z$-directional space-charge field ($E_{s|z}$) at $z=-0.04$ and $r=0$ is more negative (opposite to the applied field) in comparison to other regions (see Fig. \ref{fig:10.46ns_z}).
		Therefore, the total field ($E_{t}$) and the ionization probability at the center are smaller compared to the other regions.
		The radial component of the space-charge field ( $E_{s|r}$) plays an important role in spreading the avalanche along the transverse direction.
		The radial field at 10.46 ns at different $z$ and radius $r$ has been shown in Fig. \ref{fig:10.46ns_rad}.
		The radial field $E_{s|r}$ of ions is positive, and of electrons is negative.
		A positive sign indicates an outward radial field from the center, and a negative sign signifies an inward radial field to the center.
		It is noted that near the region $z$=-0.04 and $r$=0.005 (see Fig. \ref{fig:10.46ns_rad}) the radial field is negative, meaning the number of electrons dominates the number of ions at those places.
		The phi ($E_{s|\phi}$) component of the field measures the axial symmetry of the avalanche charge distribution.
		The charge distribution will be axially symmetric when the value of $E_{s|\phi}$ is close to zero.
		A nonzero value of $E_{s|\phi}$ signifies axial asymmetry.
		In Fig. \ref{fig:10.46ns_phi} the variation of $E_{s|\phi}$ with $z$ position and radius $r$ has been shown.
		At 10.46 ns, $E_{s|\phi}$ is not as strong as $E_{s|r}$ and $E_{s|z}$ \textcolor{black}{(maximum $E_{s|\phi}$ is only 26\% of $E_{s|z}$}).
		However, $E_{s|\phi}$ can still act as a perturbation to the applied field.
	
	The shape of the electron cloud corresponding to time 10.46 ns is shown in Fig. \ref{fig:10.46nsE}, and the magnitude of the total field ($E_{t}$) in V/cm at different voxels are shown in the color bar of the same figure.
	The maximum positive and negative deviation of the magnitude of the total field from the initially applied field are 5.62\% and 8.0\%, respectively.

\begin{figure}
	\center\subfloat[\label{fig:12.46}]{\includegraphics[scale=0.17]{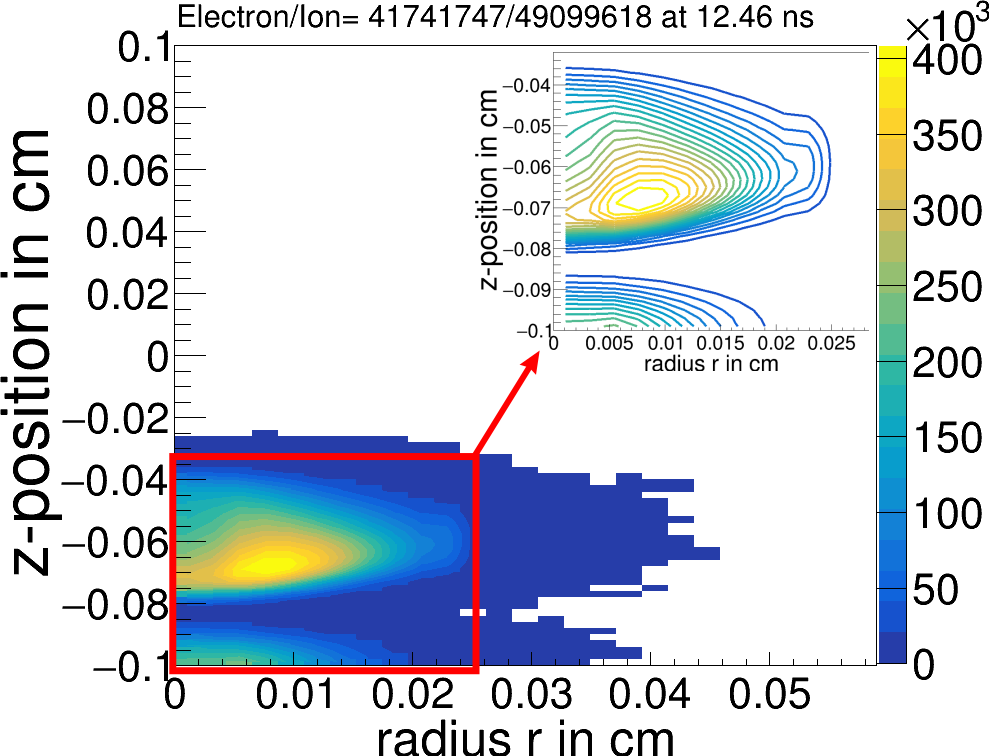}
		
	}~~\subfloat[\label{fig:12.96}]{\includegraphics[scale=0.17]{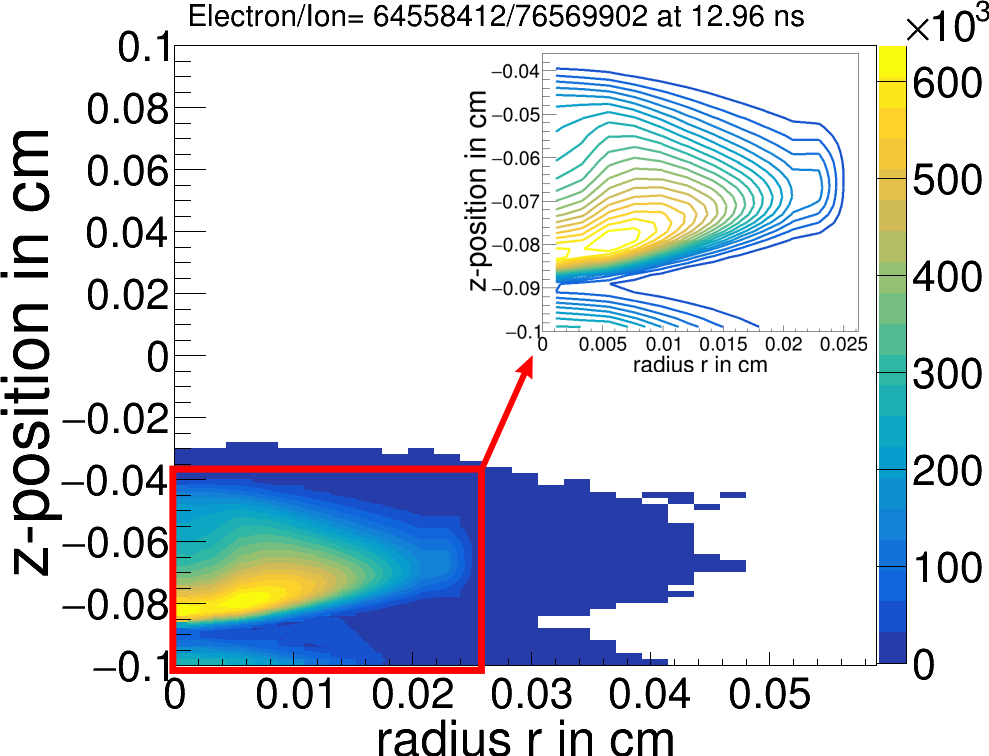}}

\center\subfloat[\label{fig:13.46}]{\includegraphics[scale=0.17]{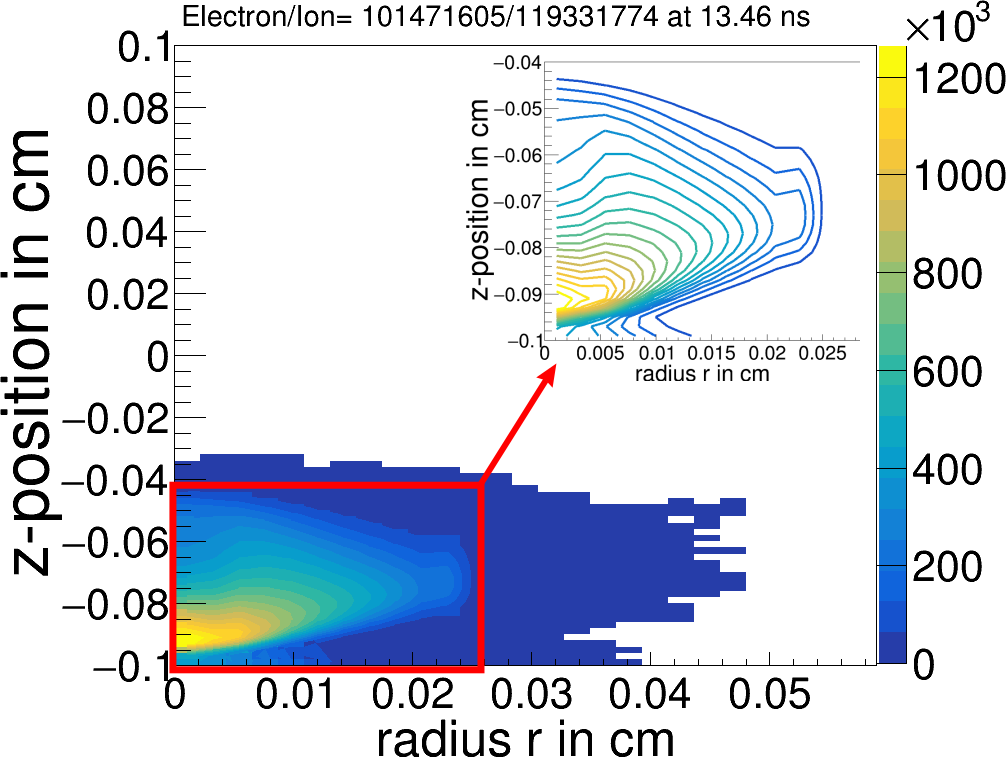}
}
	\caption{(a) Location of electrons in $z-r$ plane at time 12.46 ns, (b) Location of electrons in $z-r$ plane at time 12.96 ns, (c) Location of electrons in $z-r$ plane at time 13.46 ns.}
\end{figure}

\begin{figure}
	\center\subfloat[\label{fig:12.46ns_z}]{\includegraphics[scale=0.19]{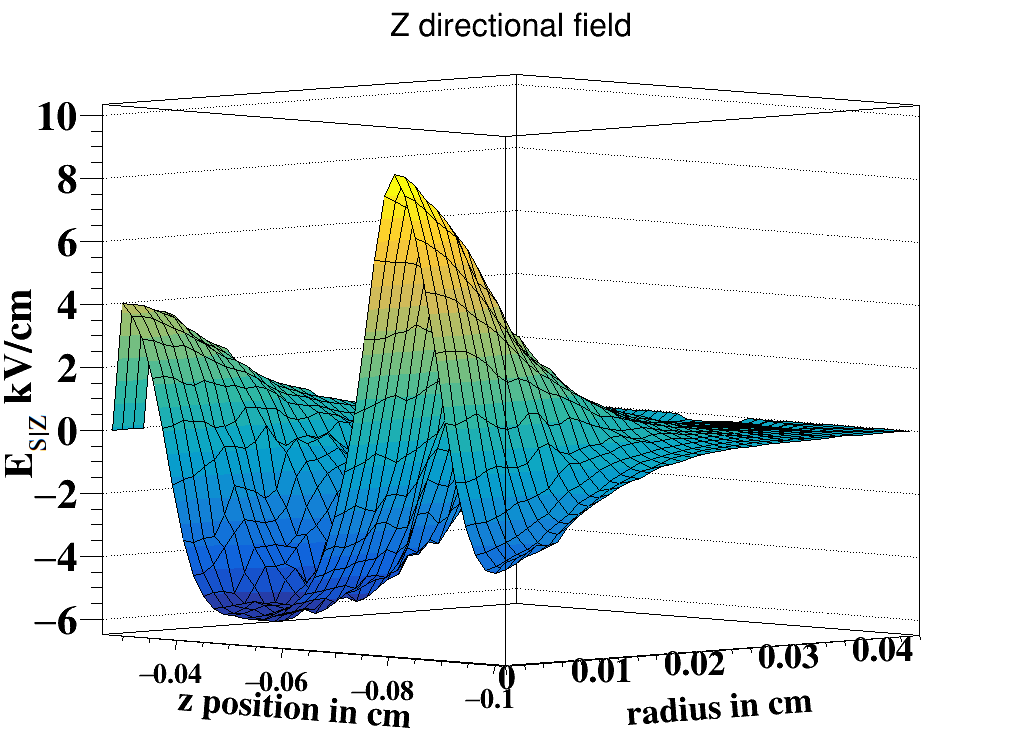}
	}\subfloat[\label{fig:12.46ns_rad}]{\includegraphics[scale=0.19]{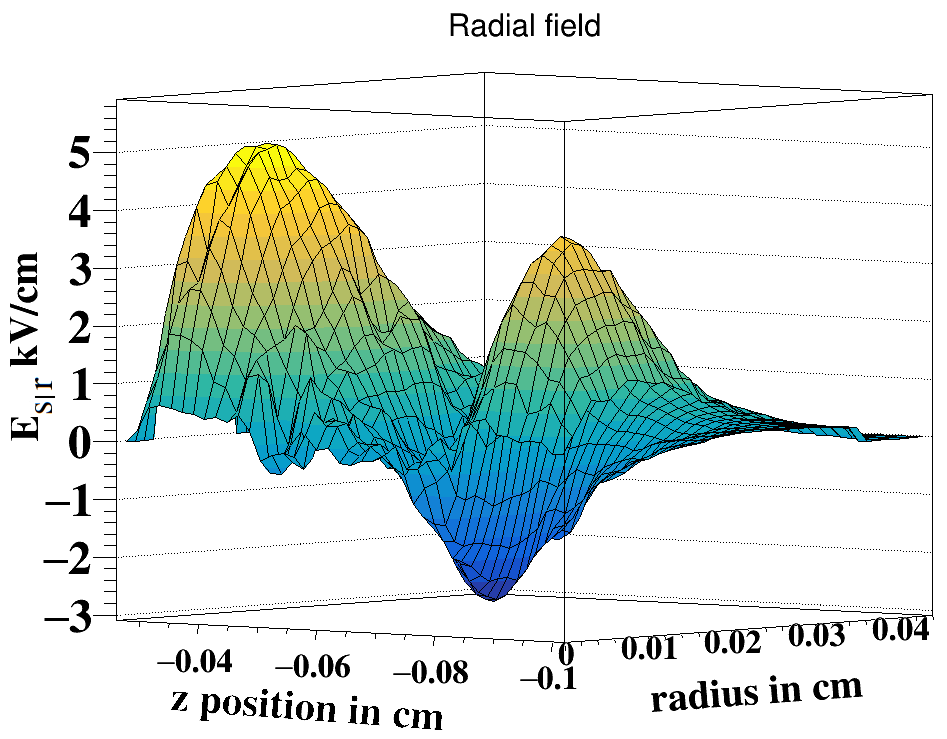}}
	
	\center\subfloat[\label{fig:12.46ns_phi}]{\includegraphics[scale=0.22]{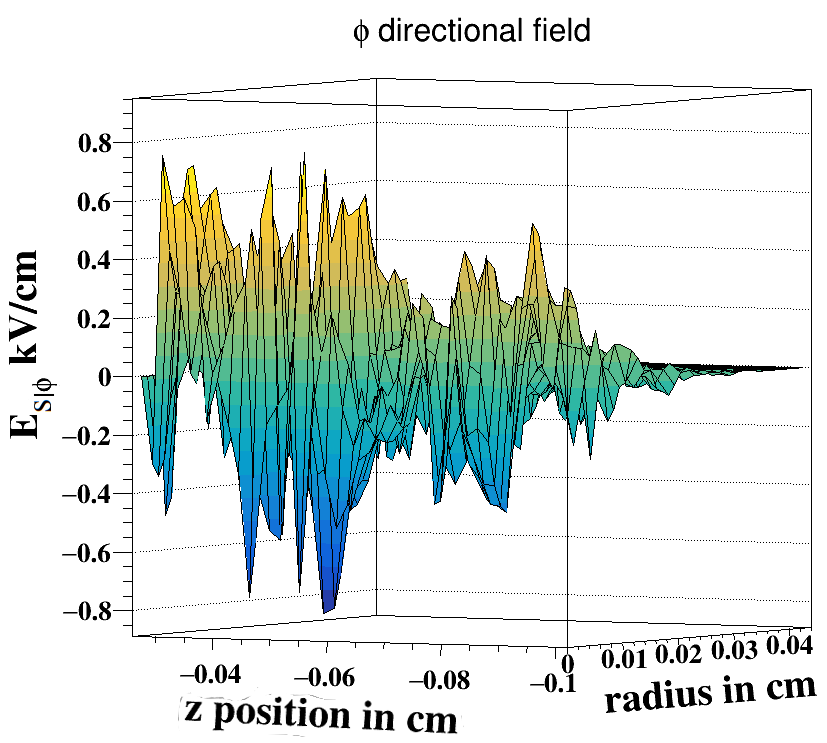}	
		
	}
	
	\caption{(a) $z$-component of space-charge field at 12.46 ns, (b) radial component of space-charge field at 12.46 ns, (c) $\phi$-component of space-charge field at 12.46 ns.}
\end{figure}	

	\item
	From the collective view of Figs. \ref{fig:12.46}, \ref{fig:12.96} and \ref{fig:13.46} it can be said that from 12.46 ns to 13.46 ns the  charge center again gradually  move back to the center $r=0$.
	The $z$-directional space-charge field ($E_{s|z}$) at 12.46 ns near $z=-0.08$ and $r=0$ becomes positive and higher than in other regions (see Figure Fig. \ref{fig:12.46ns_z}).
	Therefore, the ionization rate becomes high at that location.
	This is the reason the charge center again moves towards $r=0$.
	In Fig. \ref{fig:12.46ns_rad}, the radial field at a different location at time 12.46 ns is shown.
	At time 12.46 ns, the radial field increases significantly (absolute maximum along positive and negative directions around 5 kV/cm and 3 kV/cm, respectively) at some places, which are reflected in the transverse or radial spread ($r_{max}$=0.044 cm)  of the avalanche (see Fig. \ref{fig:12.46}).
	The variation of $E_{s|\phi}$ at different $z$ location and radius has been shown in Fig. \ref{fig:12.46ns_phi}.
	The maximum magnitude of $E_{s|\phi}$ at time 12.46 ns is 0.8 kV/cm, which is double the value of $E_{s|\phi}$ at 10.46 ns. 
	
	\par
	It is clear from Fig. \ref{fig:12.96elcden} (color bar represents the number of electrons at a given voxel) that the range of the electron cloud along the $z$ direction is -0.1 cm to -0.04 cm.
	\textcolor{black}{As shown in Fig. \ref{fig:12.96ionden}} (color bar represents the number of ions at a given voxel), the range of  the ion cloud is -0.1 cm to 0.05 cm.
	Therefore, along the $z$-axis, ions are on both sides of the electron cloud.
	Hence, at the tip and tail of the electron cloud, ions attract the nearest electrons toward them.
	\textcolor{black}{As a result, the charge distribution becomes asymmetric} along the $z$-axis.
	
	\begin{figure}
		\center\subfloat[\label{fig:12.96elcden}]{\includegraphics[scale=0.23]{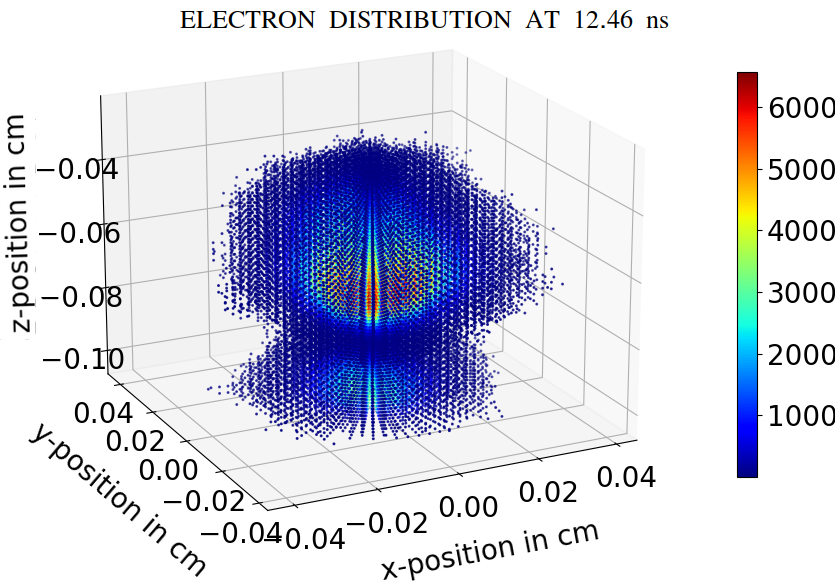}
			
		}\subfloat[\label{fig:12.96ionden}]{\includegraphics[scale=0.24]{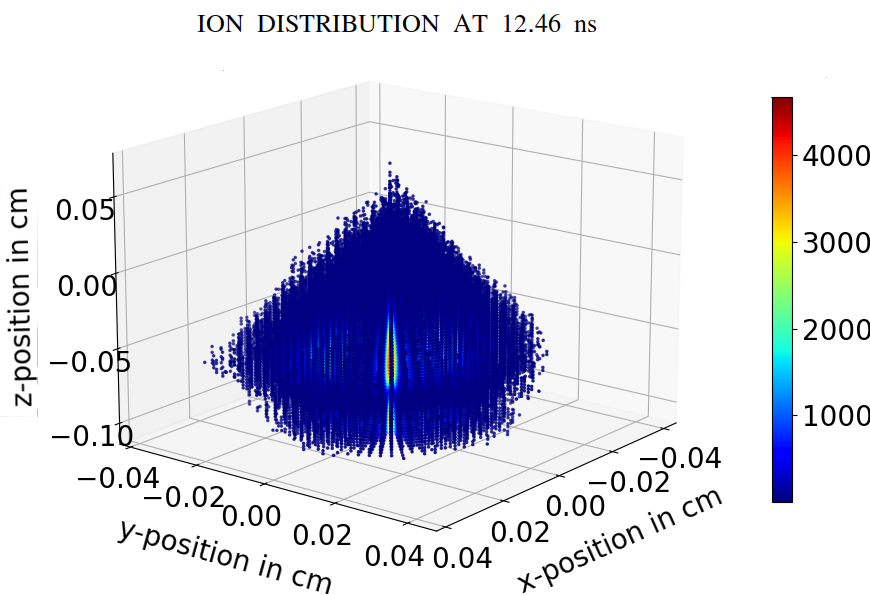}
			
		}
		\caption{(a) Shape of the electron cloud and number of electrons at different voxels (color bar) at time 12.46 ns, (b) Shape of the ion cloud and number of ions at different voxels (color bar) at time 12.46 ns.}
		
     \end{figure}	
	
	\par
	In Figs. \ref{fig:12.46nsE}, \ref{fig:12.96nsE} and \ref{fig:13.46nsE}, the shape of the electron cloud and magnitude of total field at different places of the cloud and at different times 12.46 ns,12.96 ns and 13.46 ns have been shown.
	It is found from the former figures that though the electron density at different places inside the cloud is changing with time, the overall shape of the distribution is not changed much.
	In this period of transition, the maximum positive and negative deviation of the total electric field from the applied field is 36.02\% and 27.5\%, respectively.
	
	\end{enumerate}	
	
	\begin{figure}
		\center\subfloat[\label{fig:12.46nsE}]{\includegraphics[scale=0.23]{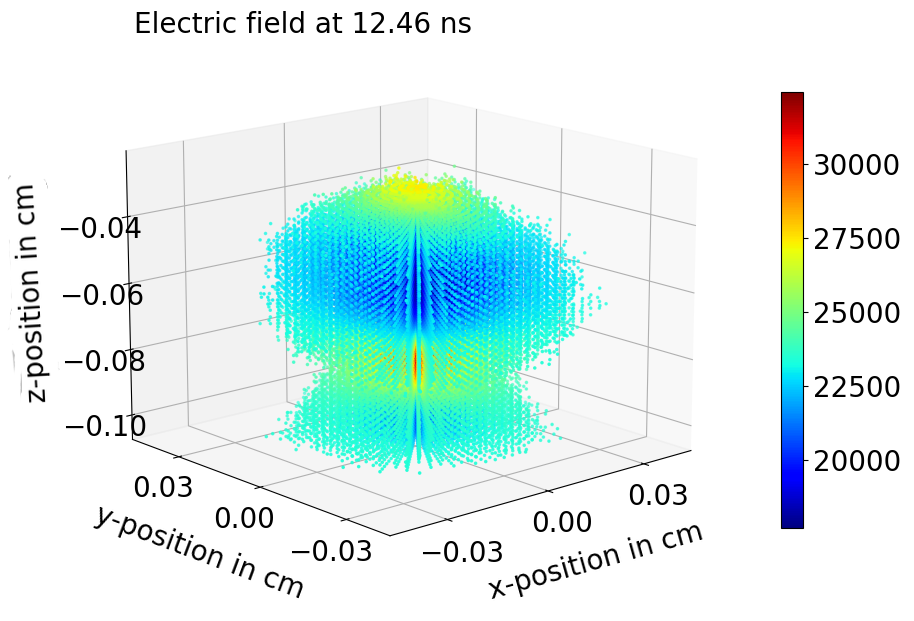}
			
		}\subfloat[\label{fig:12.96nsE}]{\includegraphics[scale=0.23]{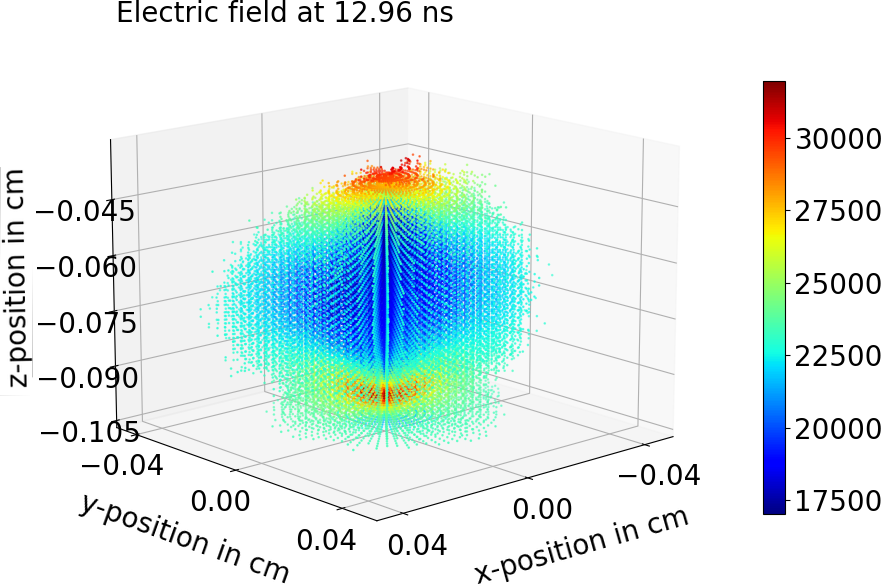}
			
		}
		
		\center\subfloat[\label{fig:13.46nsE}]{\includegraphics[scale=0.23]{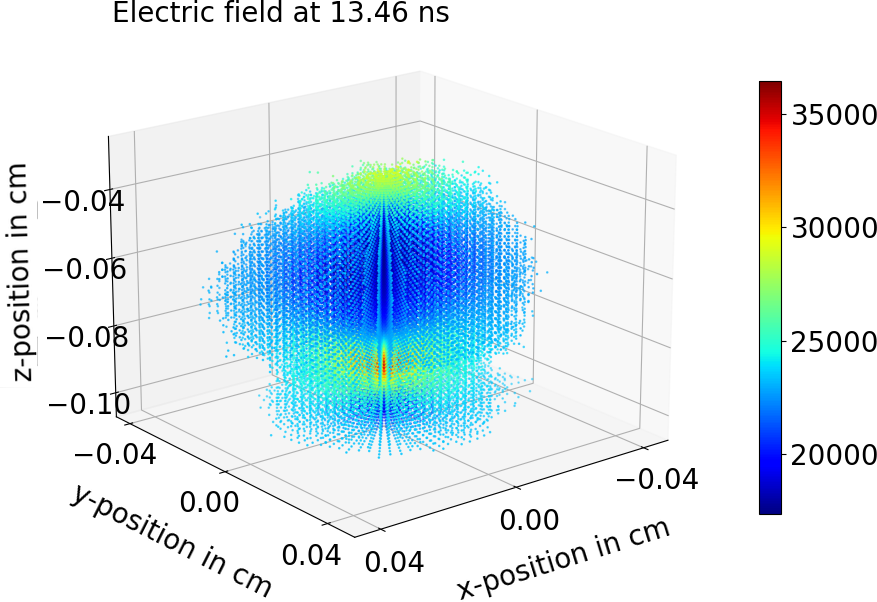}
		}
		\caption{(a) Shape of the electron cloud and electric field magnitude \textcolor{black}{($E_{t}$ \textcolor{black}{in V/cm})} at different voxels (color bar) at time 12.46 ns, (b) Shape of the electron cloud and electric field magnitude ($E_{t}$ \textcolor{black}{in V/cm}) at different voxels (color bar) at time 12.96 ns, (d) Shape of the electron cloud and electric field magnitude ($E_{t}$ \textcolor{black}{in V/cm}) at different voxels (color bar) at time 13.46 ns.} 
		\label{fig:trk_aftersp1}
		
	\end{figure}

	\subsubsection{ \textbf{After 15.46 ns, peak to saturated region of Figure Fig. \ref{fig:gain}.}}
	\begin{enumerate}[i.]
		\item At 15.46 ns, the electron gain reaches its maximum value (see Fig.  \ref{fig:gain}).
		All the electron clusters are merged together (see Fig. \ref{fig:15.46}).
		Since already many electrons have left the gas gap, the total number of ions (253,862,910) is far greater than the total number of electrons (160,516,288).
		From Fig. \ref{fig:15.46ns_z}, it is confirmed that due to negative $E_{s|z}$, the net z-directional field has reduced by a large amount at the tip and the center ($r=0$) of the avalanche.
		Also, due to the significant difference in the electron and ion numbers, the radial field of the ions dominates; hence, only a positive radial field exists everywhere.
		As a result, the remaining electrons feel a strong radial force towards the center, as shown in Fig. \ref{fig:15.46ns_rad}.
		Therefore, the transverse spread is stopped, and all the electrons tend to converge toward the center.
		The component $E_{s|\phi}$ attains a maximum value of 2 kV/cm approximately (see Fig. \ref{fig:15.46ns_phi}) which indicates significant axial asymmetry of the \textcolor{black}{space-charge distribution}. 
		In Fig. \ref{fig:15.46nsE}, the shape of the electron cloud and electric field at different  voxels has been shown. 
		The maximum  positive and negative deviation of a total electric field from the applied field at 15.46 ns is 84.4\% and 50.5\%, respectively.

	\begin{figure}
		\center\subfloat[\label{fig:15.46}]{\includegraphics[scale=0.17]{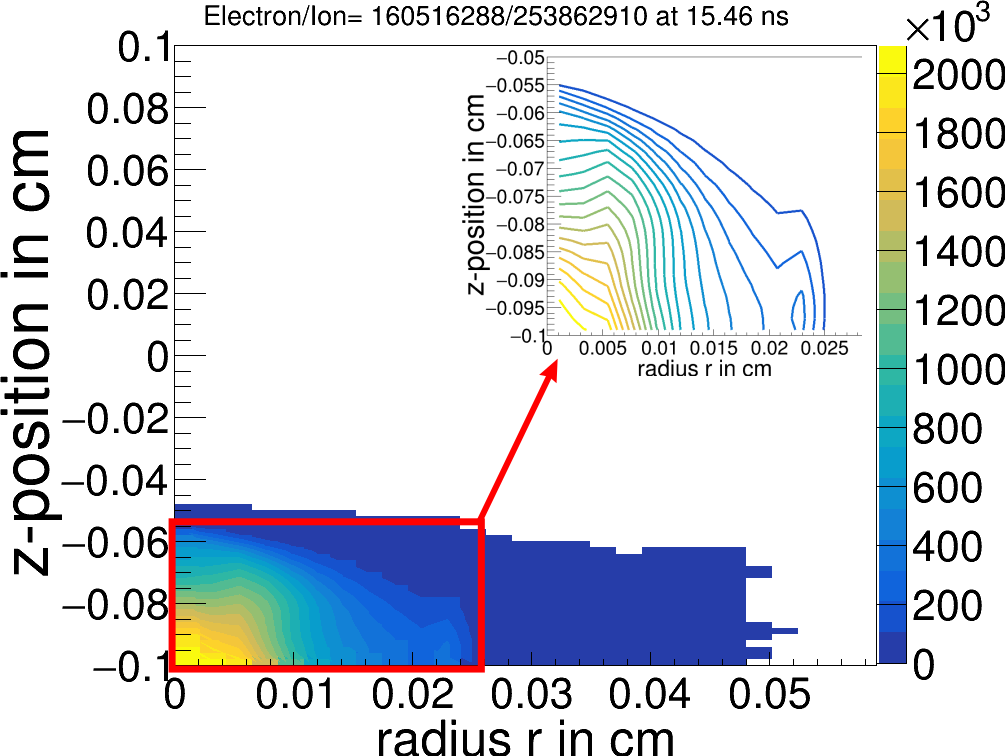}
			
		}~~\subfloat[\label{fig:15.46nsE}]{\includegraphics[scale=0.23]{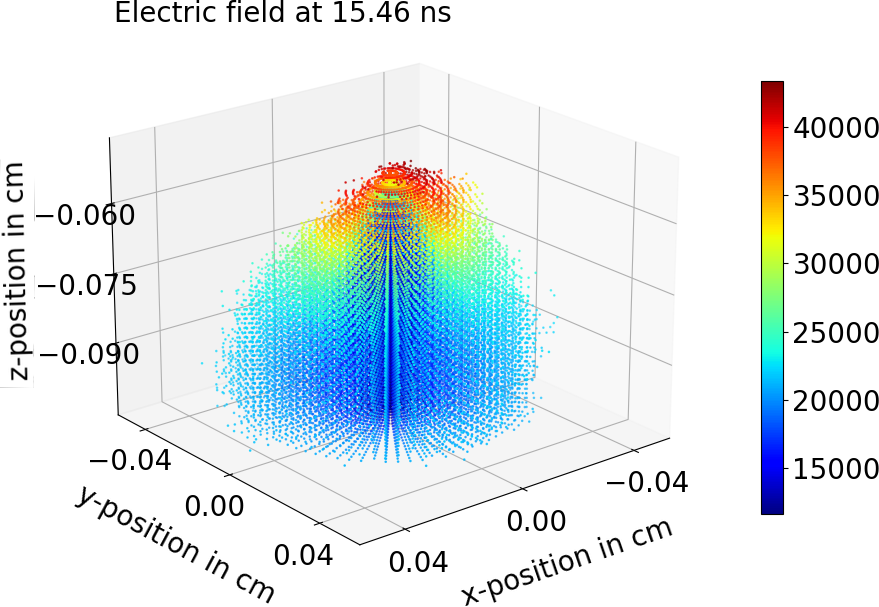}
		}

			\center\subfloat[\label{fig:16.96}]{\includegraphics[scale=0.17]{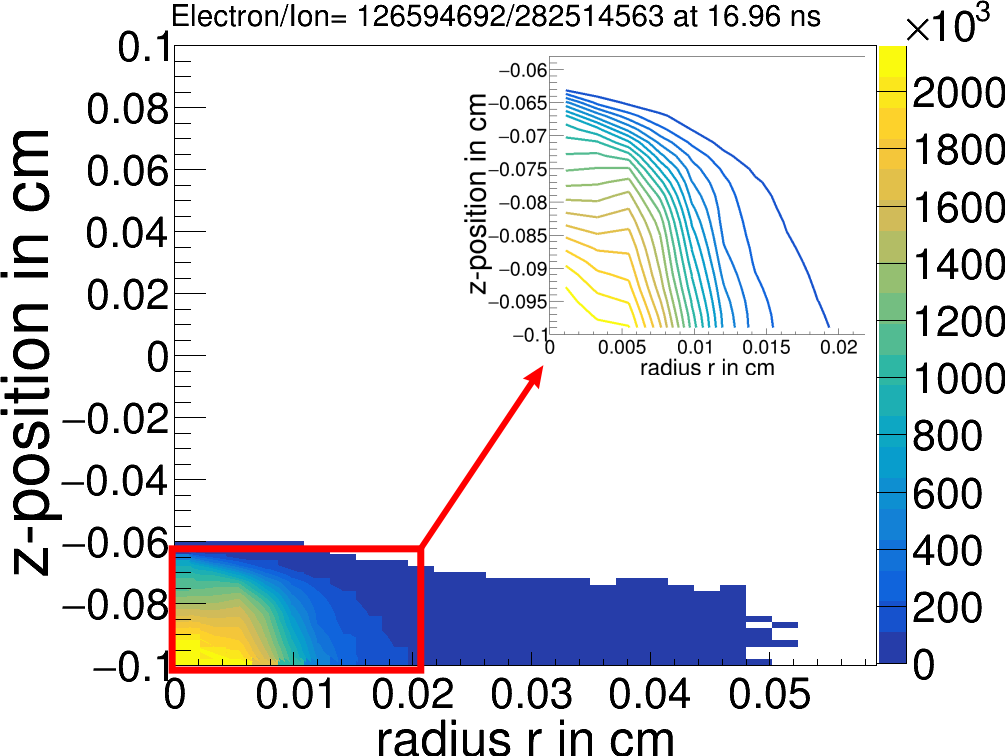}
			
		}~~\subfloat[\label{fig:16.96nsE}]{\includegraphics[scale=0.23]{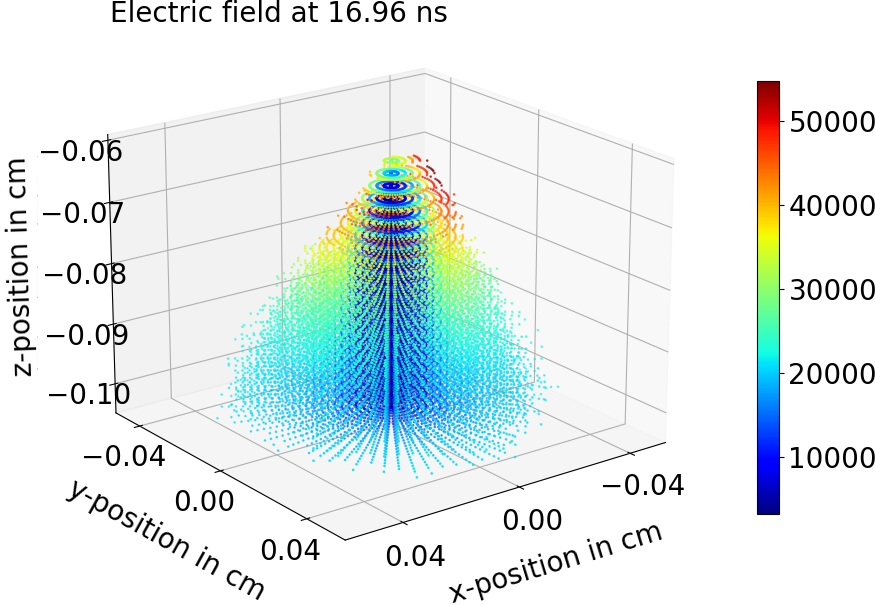}
			
		}
		\caption{(a) Location of electrons in $z-r$ plane at time 15.46 ns, (b) Shape of the electron cloud and electric field magnitude ($E_{t}$ \textcolor{black}{in V/cm}) at different voxels (color bar) at time 15.46 ns, (c) Location of electrons in $z-r$ plane at time 16.96 ns, (d) Shape of the electron cloud and electric field magnitude ($E_{t}$ \textcolor{black}{in V/cm}) at different voxels (color bar) at time 16.96 ns.}\label{fig:trk_aftersp2}
		
	\end{figure}
      
	\begin{figure}
		\center\subfloat[\label{fig:15.46ns_z}]{\includegraphics[scale=0.19]{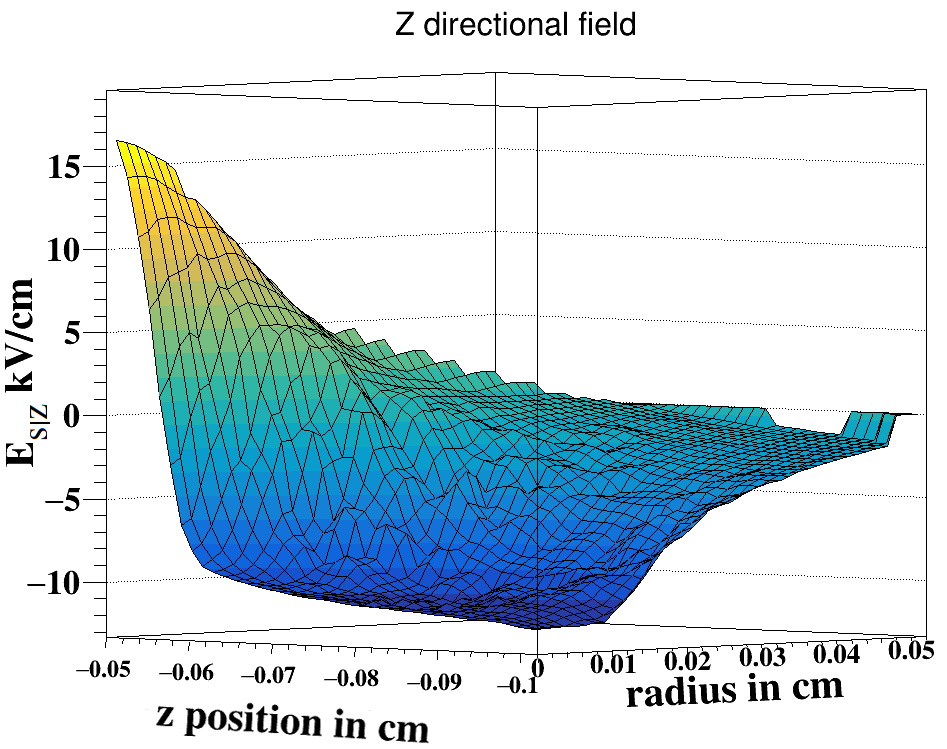}
		}\subfloat[\label{fig:15.46ns_rad}]{\includegraphics[scale=0.19]{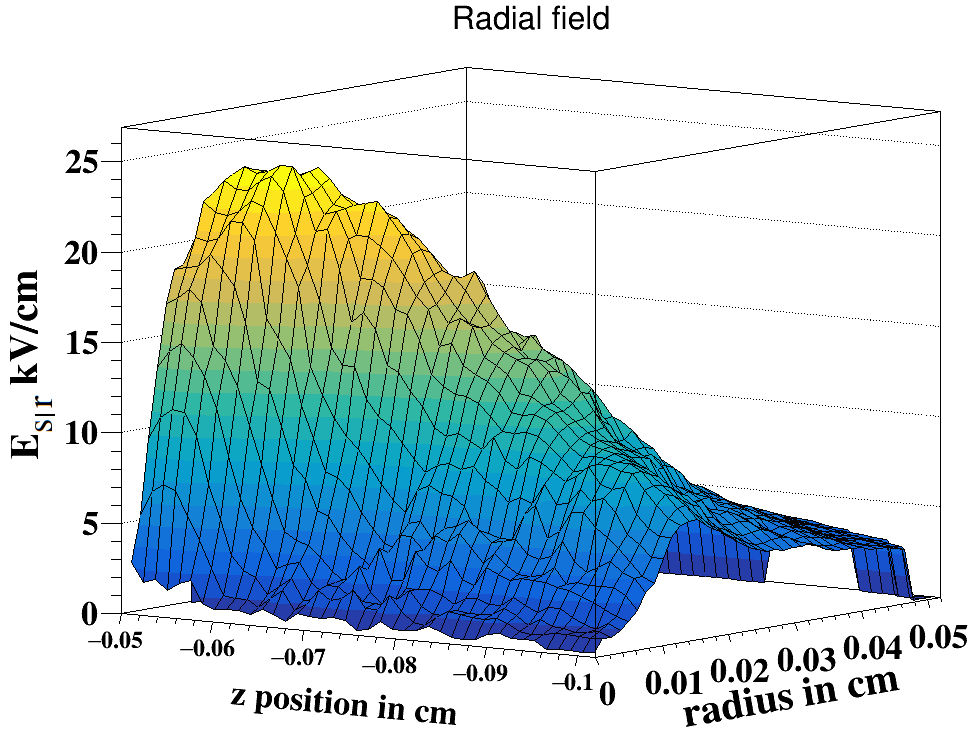}}
		
		\center\subfloat[\label{fig:15.46ns_phi}]{\includegraphics[scale=0.22]{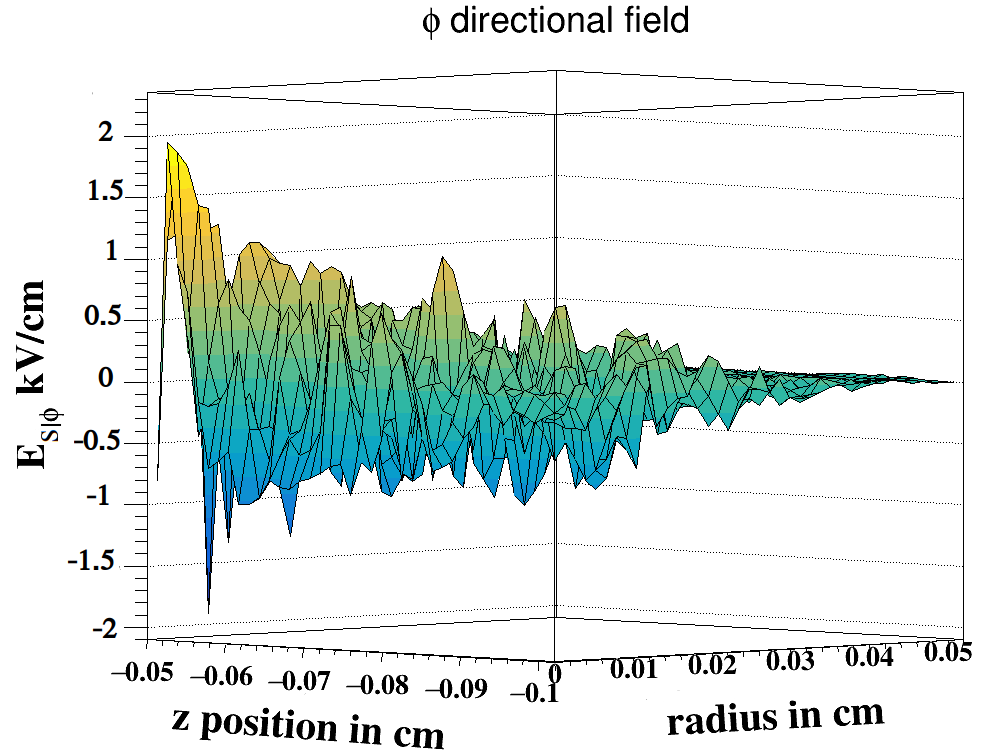}	
			
		}
		
		\caption{(a) $z$-component of space-charge field at 15.46 ns, (b) radial component of space-charge field at 15.46 ns, (c) $\phi$-component of space-charge field at 15.46 ns.}
	\end{figure}
	
	\begin{figure}
		\center\subfloat[\label{fig:18.46}]{\includegraphics[scale=0.17]{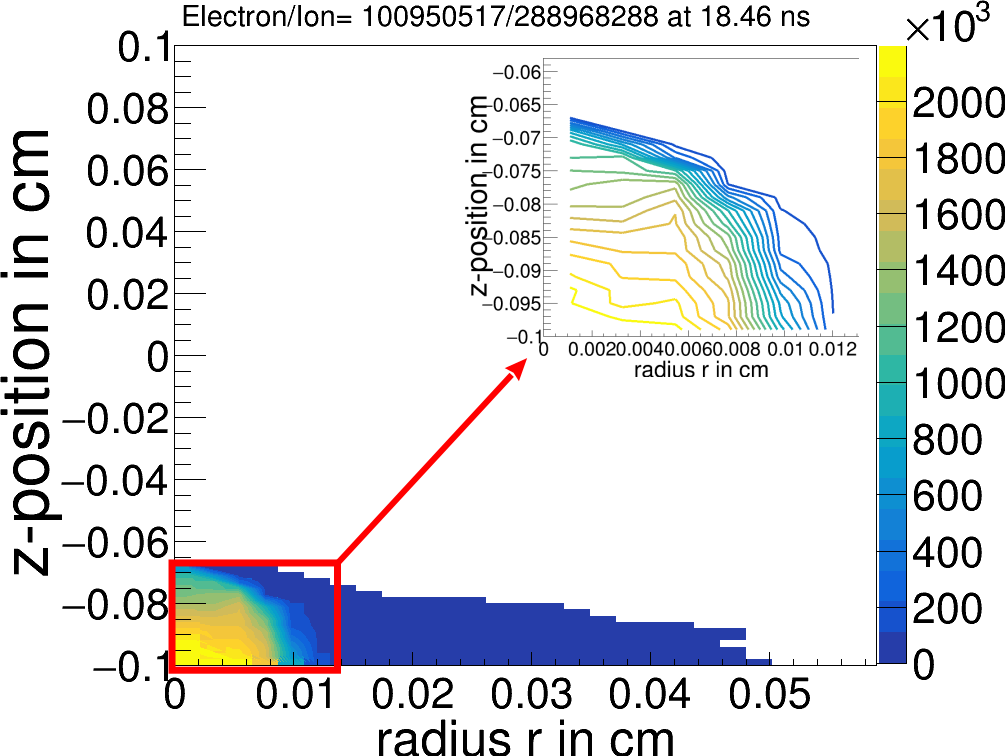}
			
		}~~\subfloat[\label{fig:18.96nsE}]{\includegraphics[scale=0.23]{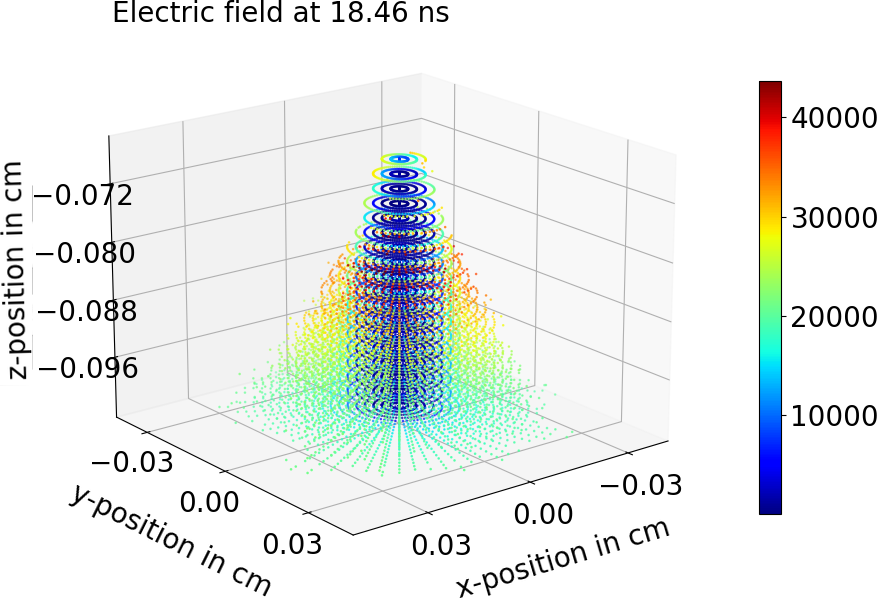}
			
		}
		
			\center\subfloat[\label{fig:23.46}]{\includegraphics[scale=0.17]{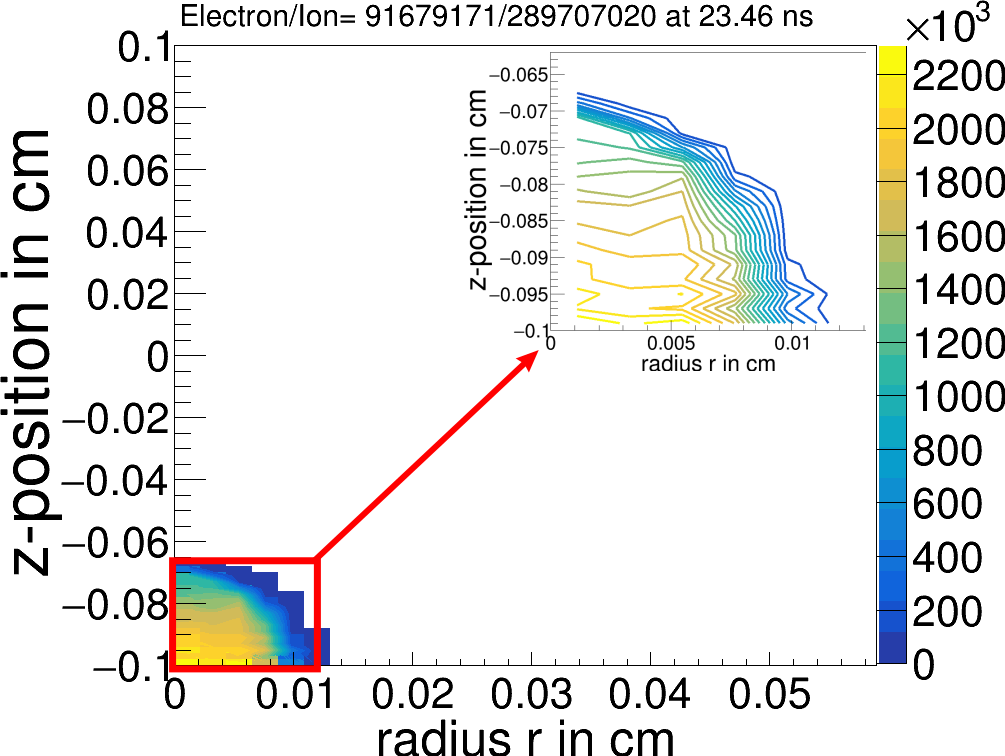}
		}~~\subfloat[\label{fig:23.46nsE}]{\includegraphics[scale=0.23]{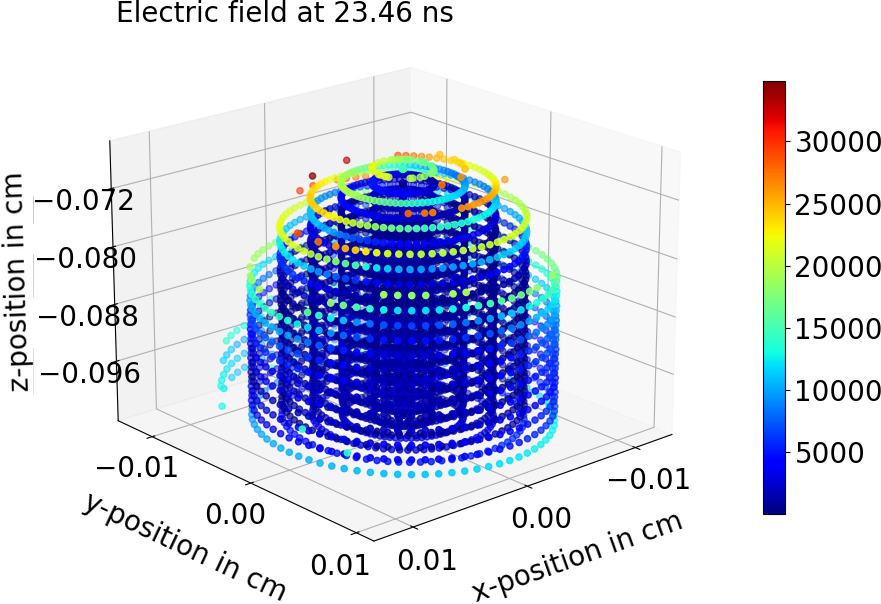}
		}
		
		\caption{(a) Location of electrons in $z-r$ plane at time 18.46 ns, (b) Shape of the electron cloud and electric field magnitude ($E_{t}$ \textcolor{black}{in V/cm}) (color bar) at different voxels at time 18.46 ns, (c) Location of electrons in $z-r$ plane at time 23.46 ns, (d) Shape of the electron cloud and electric field magnitude ($E_{t}$ \textcolor{black}{in V/cm}) (color bar) at different voxels at time 23.46 ns.}\label{fig:trk_aftersp2}
		
	\end{figure}
	
	\item
	At time 16.96 ns, there are 126,594,692 electrons and 282,514,583 ions left inside the gas gap (see Fig. \ref{fig:16.96}).
	This is the time when the space-charge field attains its maximum value.
	The maximum  positive and negative deviation of the magnitude of the total field (space-charge + applied field) from the initially applied field are 132.8\% and 86.2\% respectively.
	The shape of the electron cloud becomes conical (see Fig. \ref{fig:16.96nsE}), where at the tip, the field is reduced drastically, and at the tail, it becomes more than double the initial value of the field. 
		
		\item 
		At 18.46 ns, the gain curve with space-charge effect of Fig. \ref{fig:gain} contains a second knee point after which the saturation region starts.
		At this stage, there are 100,950,517 electrons and 288,968,288 ions.
		The density contours are not smooth near the center (see Fig. \ref{fig:18.46}).
		There are three possibilities for the reduction of the number of outermost electrons  (i) attachment, (ii) leaving the gas gap, and (iii) at some places, the inward radial field is much stronger so that it attracts the electrons towards the center.
		Therefore, the conical shape gradually modifies into cylindrical (see Fig. \ref{fig:18.96nsE}).
		The maximum positive and negative deviation of the magnitude of the total field ($E_{t}$) from the initially applied field are 85.8\% and 99.9\%.
		
		\item
		At 23.46 ns, the gain of the avalanche is saturated.
		In most of the voxels, the total field is less than 1 kV/cm (see Fig. \ref{fig:23.46nsE}).
		Therefore, at those places, the drift velocity of the electrons is less than 0.008 cm/ns (see Fig. \ref{fig:23.46nsdrft}).
		Also, in these low-field regions, the attachment coefficient ($\eta$) dominates over the first Townsend coefficient (alpha) ($\alpha$). Hence the value of $\alpha - \eta$ becomes negative (see Fig. \ref{fig:23.46nsalpEta}).
		At higher field regions, there are some electrons capable of producing ionisations.
		However, the number of electrons created in ionisations is compensated by the number of electrons attached; that's why the electron gain saturates, as shown in Fig. \ref{fig:gain}.
		The maximum positive and negative deviation of the magnitude of the total field ($E_{t}$) from the initially applied field are 41.5\% and 99.8\%.
		\end{enumerate}	

	\begin{figure}
		\center\subfloat[\label{fig:23.46nsdrft}]{\includegraphics[scale=0.23]{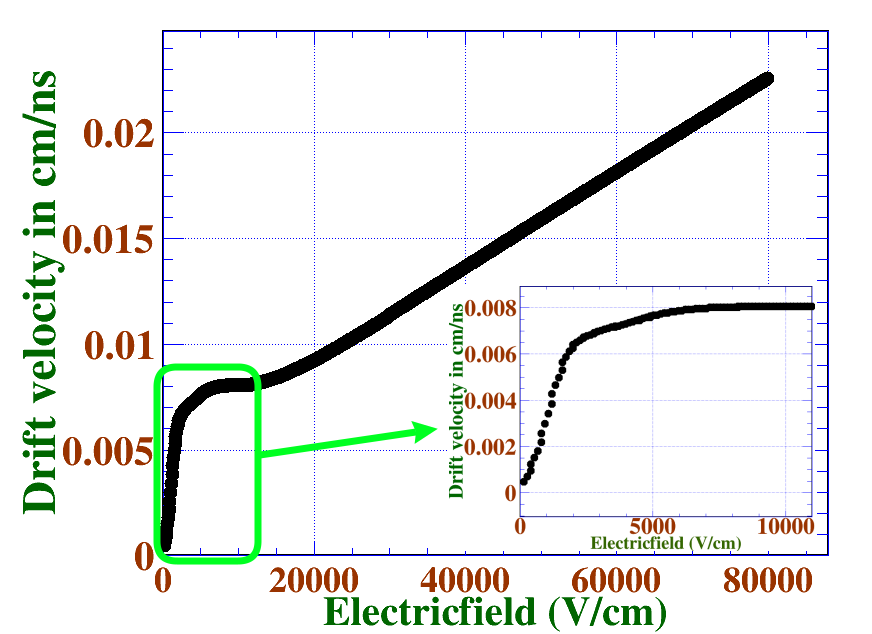}
			
		}\subfloat[\label{fig:23.46nsalpEta}]{\includegraphics[scale=0.23]{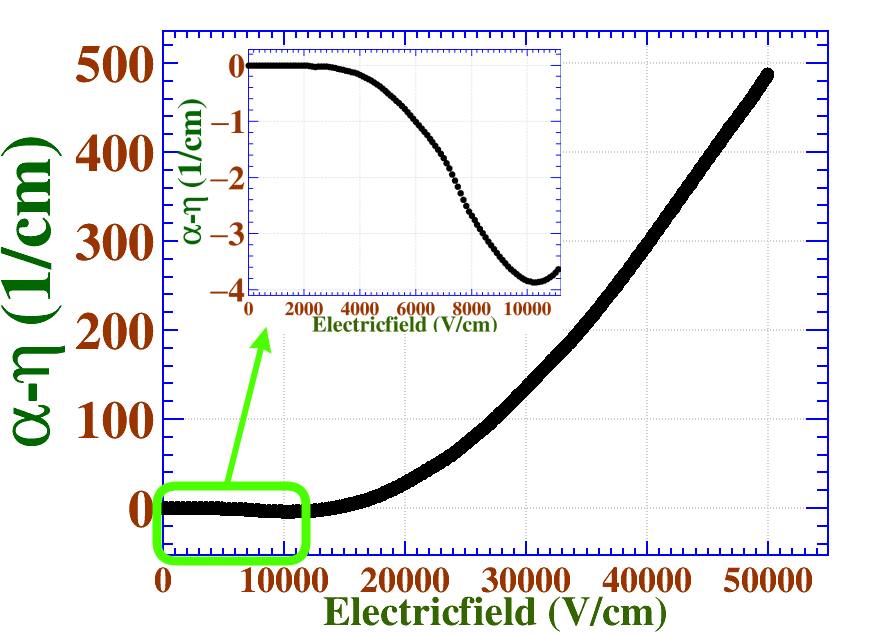}
			
		}
		
		\caption{(a) Drift velocity as a function of electric field, (b) Effective Townsend coefficient as a function of electric field. \label{fig:driftvel}}
		
	\end{figure}

 \section{
Avalanche simulations inside the RPC with $\ce{C_2H_2F_4}$, $\ce{i-C_4H_{10}}$, and $\ce{SF_6}$ gas mixture}	\label{sec:examplewithsf6}
 \subsection{\textcolor{black}{\textbf{Avalanche in absence of negative ion}}}\label{subsec:aval_conv_gas_nonegion}
 \begin{figure}[H]
 	\center\subfloat[\label{fig:gain_sf6_nosp}]{\includegraphics[width=.55\linewidth]{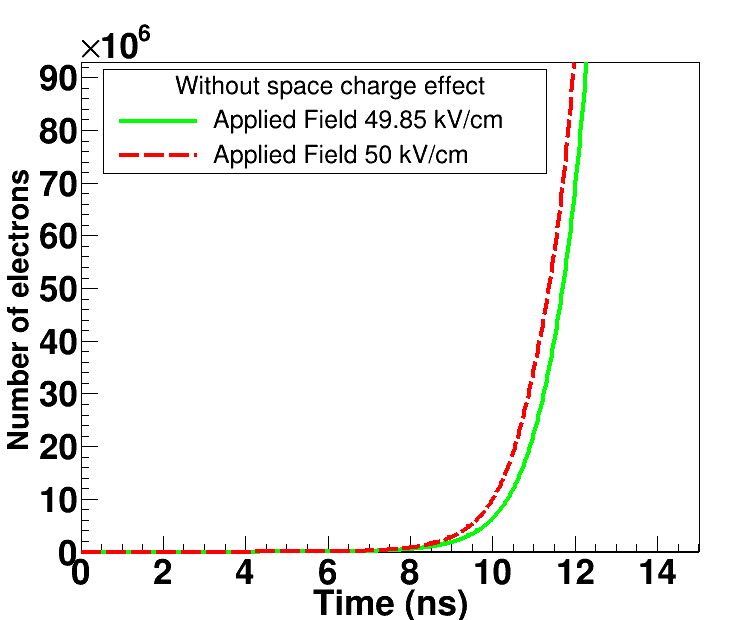}}\subfloat[\label{fig:gain_sf6_wsp}]{\includegraphics[width=.55\linewidth]{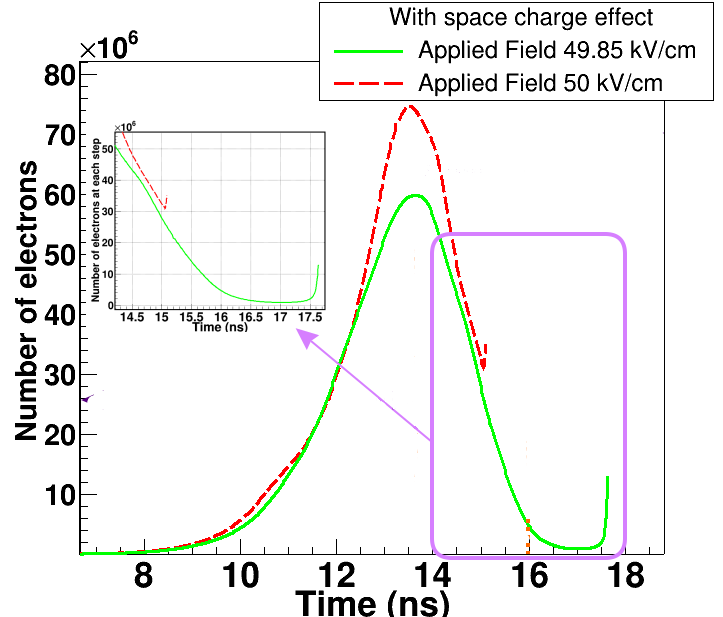}}
 	
 	\subfloat[\label{fig:Elc_field_t_sf6}]{\includegraphics[width=.53\linewidth]{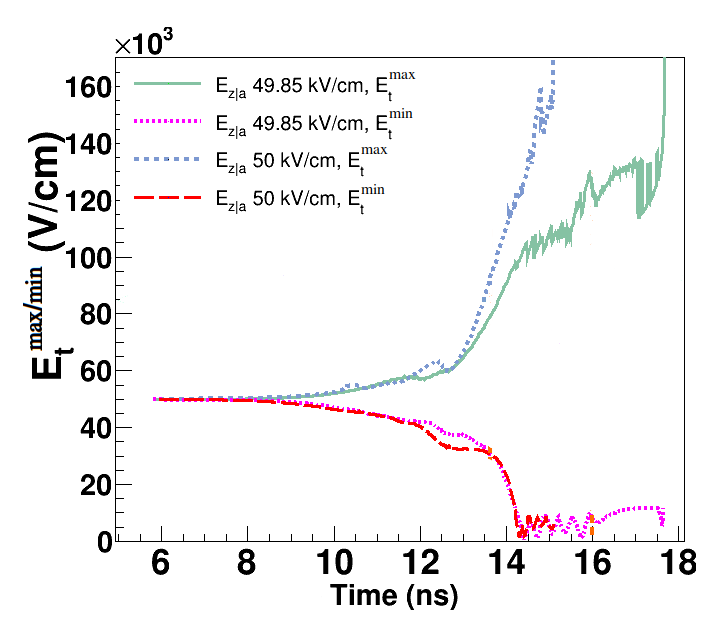}}
 	
 	\caption{Variation of electron gain with time steps of the avalanche for electric fields of 49.85 kV/cm, and 50 kV/cm, (a) without space-charge effect, and (b) with space-charge effect, (c) variation of maximum and minimum values of total electric field with time. \label{fig:gain_comp_sf6}}
 	
 \end{figure}
\textcolor{black}{In this particular section, we present another example of the avalanches occurring inside an RPC with a 2 mm gas-gap. For our investigation, we used a gas-mixture containing 97\% $\ce{C_2H_2F_4}$, 2.5\% $\ce{i-C_4H_{10}}$, and 0.5\% $\ce{SF_6}$. A muon track and primary ionisations inside the gas gap are simulated using HEED.
We performed avalanche simulations, first without taking into account the space-charge effect, and then with the space-charge effect. Two different electric fields 49.85 kV/cm, and 50 kV/cm along +z direction are considerd for these simulations. Our findings show the variation in the number of electrons with time, as shown in Figs. \ref{fig:gain_sf6_nosp} (without space-charge effect) and \ref{fig:gain_sf6_wsp} (with space-charge effect). From Fig. \ref{fig:gain_sf6_nosp} it is found that in absence of the space-charge effect the number of electrons grow continuously with \textcolor{black}{time. Similar} to the case discussed in Section \ref{sec:5_instanceAvalancheConst}, the number of electrons in the presence of the space-charge effect initially grows with time, reaches a peak, and then decreases, as shown in Fig. \ref{fig:gain_sf6_wsp}. In the same figure, the green curve exhibits a brief saturation region after 16.5 ns, followed by a steep increase at 17.5 ns. In the case of an applied field of 50 kV/cm (red curve in Fig. \ref{fig:gain_sf6_wsp}), a saturation region is not observed. However, in this case, the number of electrons reaches a peak, starts decreasing, and suddenly increases without having reached a saturation period. The variation of the maximum and minimum of the total field $E_t^{max/min}$ with time for both applied fields is shown in Fig. \ref{fig:Elc_field_t_sf6}. }
\subsection{\textcolor{black}{\textbf{Avalanche in presence of negative ion}}}
  \begin{figure}[H]
	\center\subfloat[\label{fig:gain_negion_sf6}]{\includegraphics[width=.55\linewidth]{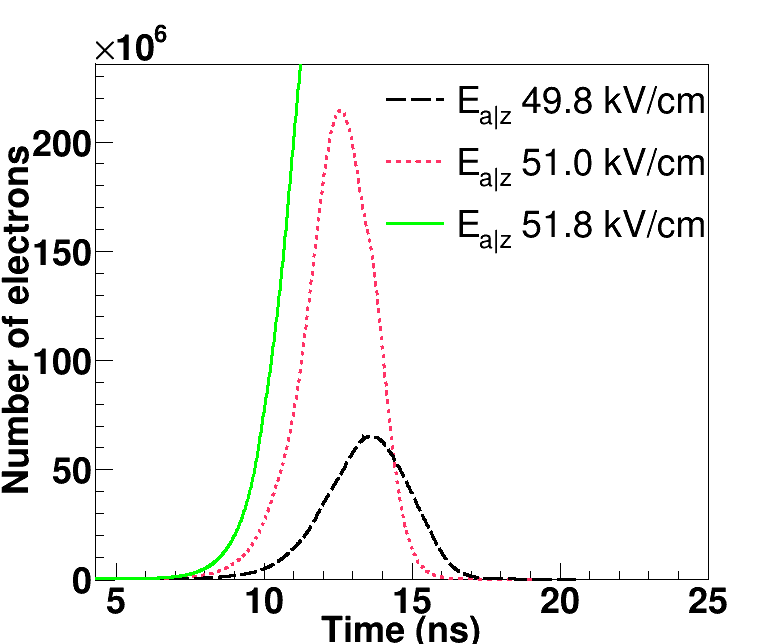}} \subfloat[\label{fig:etmax_negion_sf6}]{\includegraphics[width=.5\linewidth]{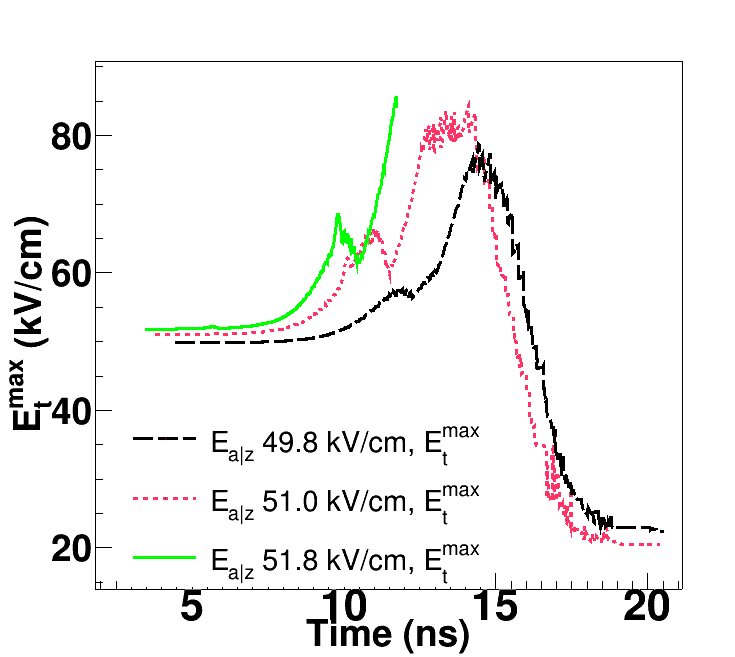}}

\subfloat[\label{fig:etmin_negion_sf6}]{\includegraphics[width=.515\linewidth]{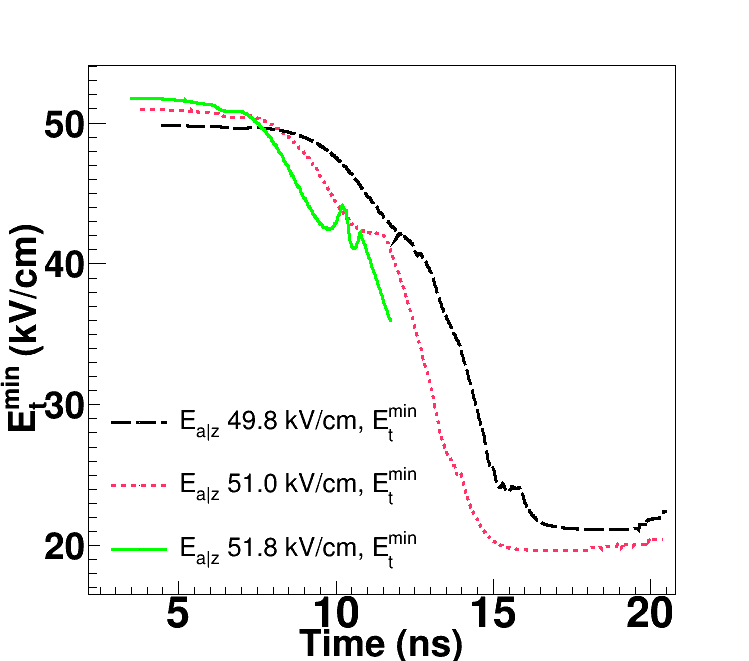}}\subfloat[\label{fig:ratio_negion_sf6}]{\includegraphics[width=.54\linewidth]{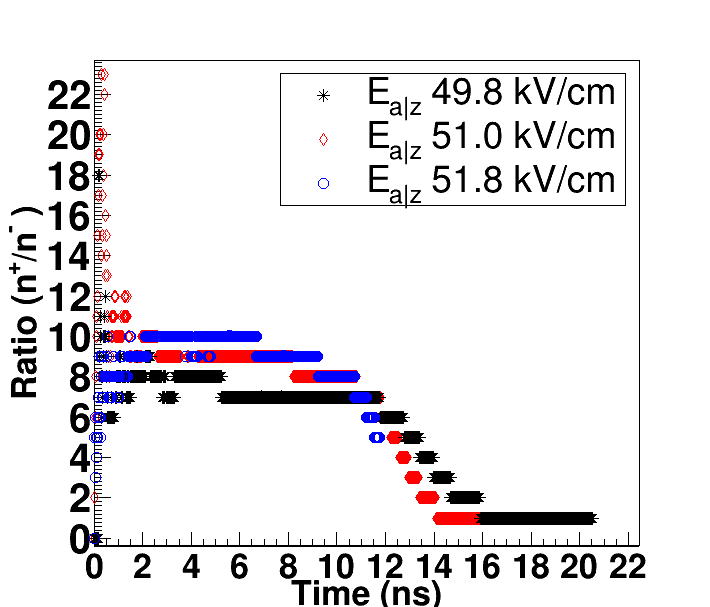}}

	\caption{(a) Variation of number of electrons with time in presence of negative ions. Variation of total electric field, (b) $E_{t}^{max}$ and (c) $E_{t}^{min}$ with time. (d) Variation of ratio between number of positive ions ($n^{+}$) and  negative ions ($n^{-}$) with time for a gas mixture  97\% of $\ce{C_2H_2F_4}$, 2.5\% of $\ce{i-C_4H_{10}}$, and 0.5\% of $\ce{SF_6}$.}
\end{figure}

\textcolor{black}{Avalanche processes take place within the RPC with the space charge effect, under three distinct applied electric fields: 49.8 kV/cm, 51 kV/cm, and 51.8 kV/cm. The geometric configuration and gas mixture remain consistent with those described in subsection \ref{subsec:aval_conv_gas_nonegion}. The variation in the number of electrons over time for the three different applied electric fields is depicted in Fig. \ref{fig:gain_negion_sf6}. As the applied electric field increases, the peaks of the curves in Fig. \ref{fig:gain_negion_sf6} shift towards smaller time values (from 13.5 ns at 49.8 kV/cm to 12.5 ns at 51 kV/cm). At 51.8 kV/cm, the number of electrons exhibits a significant increase beyond 10 ns. However, unlike that observed in subsection \ref{subsec:aval_conv_gas_nonegion}, no saturation region or abrupt steep rise in electron count is observed here. The fluctuations in $E^t_{max}$ and $E^t_{min}$ corresponding to the three applied voltages are illustrated in Figs. \ref{fig:etmax_negion_sf6} and \ref{fig:etmin_negion_sf6}, respectively. With the growth of electrons, positive and negative ions, $E^t_{max}$ surpasses the applied field, while $E^t_{min}$ falls below it. Following the peak (refer to the red and black lines in Fig. \ref{fig:gain_negion_sf6}), the electron count begins to decline. Consequently, for the applied fields of 49.85 kV/cm and 51 kV/cm (see Figs. \ref{fig:etmax_negion_sf6} and \ref{fig:etmin_negion_sf6}), $E^t_{max}$ decreases and eventually becomes nearly equivalent to $E^t_{min}$.  Nonetheless, for the scenario at 51.8 kV/cm, the electron count never decreases, maintaining an upward trend in $E^t_{max}$ and $E^t_{min}$, exhibiting an opposite trend as shown in Figs. \ref{fig:etmax_negion_sf6} and \ref{fig:etmin_negion_sf6}.}
	\par\textcolor{black}{ The time variation of the ratio between the number of positive ions ($n^{+}$) and negative ions ($n^{-}$) is depicted in Fig. \ref{fig:ratio_negion_sf6}. The presence of SF$_6$ enhances the electronegativity of the gas mixture used here, resulting in a high attachment coefficient. Consequently, the probability of negative ion generation is high. From Fig. \ref{fig:ratio_negion_sf6}, it can be concluded that for an applied field of 49.8 kV/cm at the initial stage of the avalanche, there are more positive ions than negative ions. This leads to a maximum ratio between $n^{+}$ and $n^{-}$ exceeding 22. As time progresses, this ratio gradually decreases and approaches 1 at the end of the avalanche. The same trend is observed for the case of an applied field of 51 kV/cm. However, when the applied field is 51.8 kV/cm, the avalanche diverges, and at 12 ns, the final ratio is approximately 6.	
		 }

	\section{Timing performance of pAvalancheMC using OpenMp}
	\label{sec:6_speedUp}
	We have used the OpenMP multithreading technique to parallelize the class pAvalancheMC of Garfield++.
	In the following, a timing RPC of area 30 cm $\times$ 30 cm, 0.3 mm gas gap, and 2 mm thick bakelite electrode has been studied in order to assess the timing performance of the parallelized code.
	A voltage of $\pm$1720 V  has been applied on the graphite surface.
	As a result, the average electric field inside the gas gap is 43 kV/cm. A gas mixture of $\ce{C_2H_2F_4}$ (85\%), $\ce{i-C_4H_10}$ (5\%), $\ce{SF_6}$ (10\%) is used. 
	\par
\begin{figure}
		\center\subfloat[\label{fig:timePer_nosp}]{\includegraphics[width=.5\linewidth]{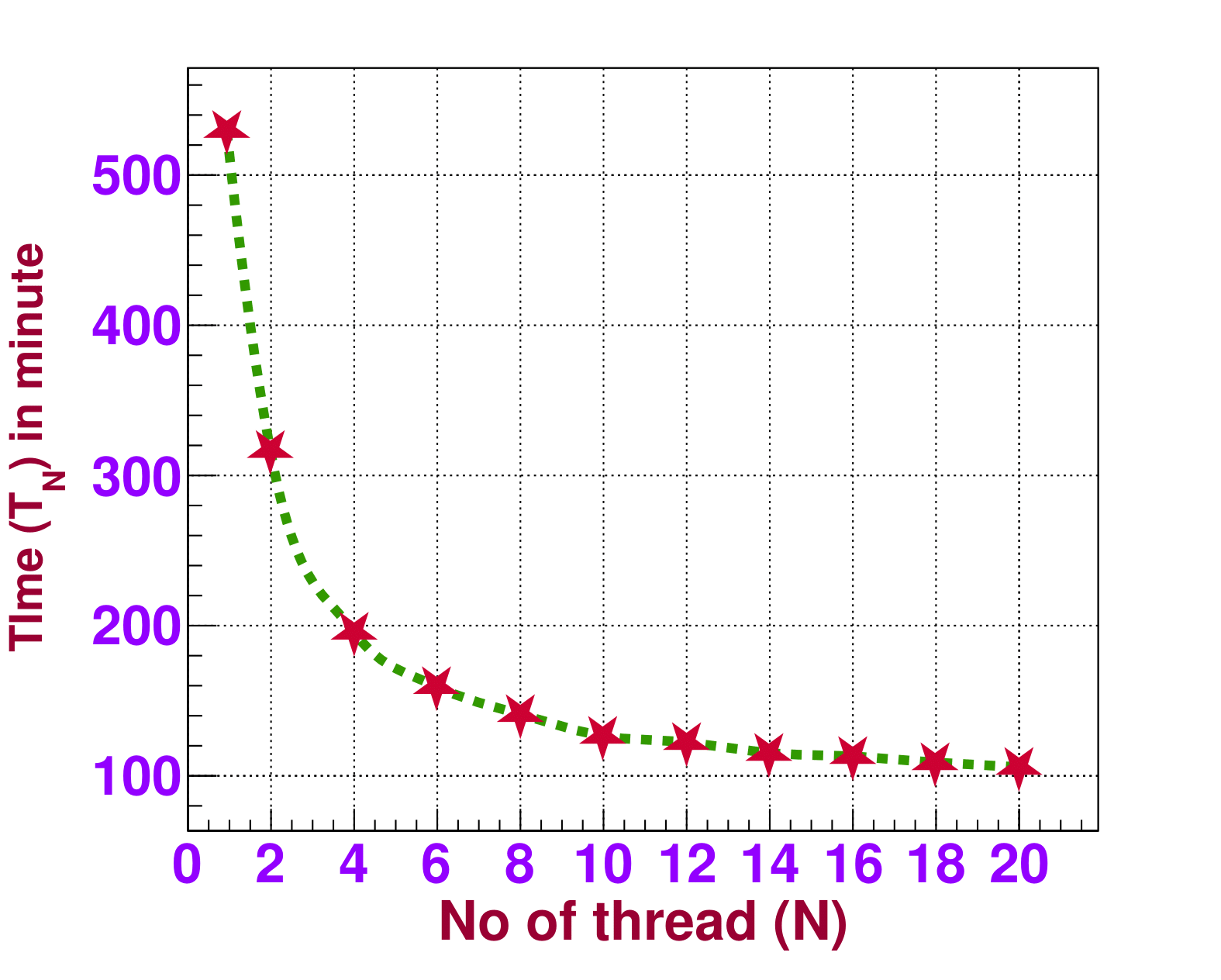}}\subfloat[\label{fig:speedUp_nosp}]{\includegraphics[width=.5\linewidth]{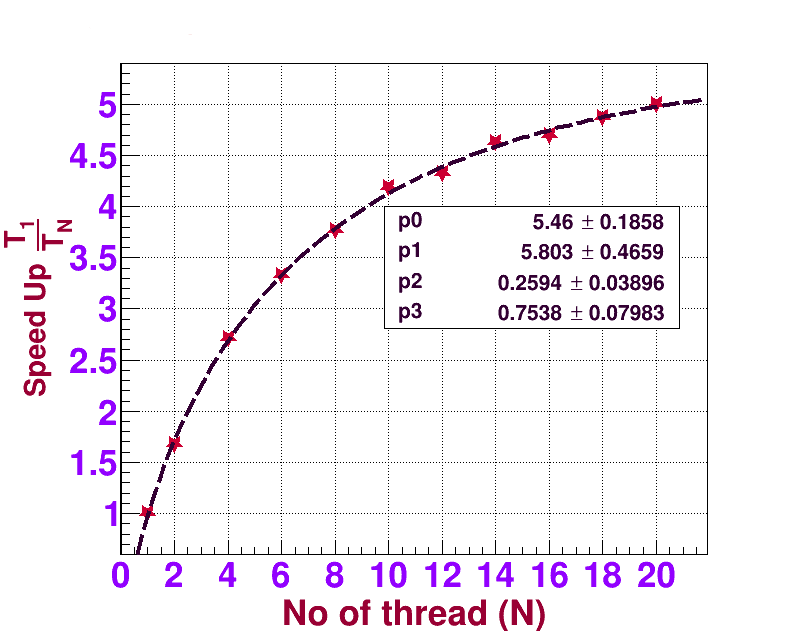}}
		
		\center\subfloat[\label{fig:timePer_wsp}]{\includegraphics[width=.5\linewidth]{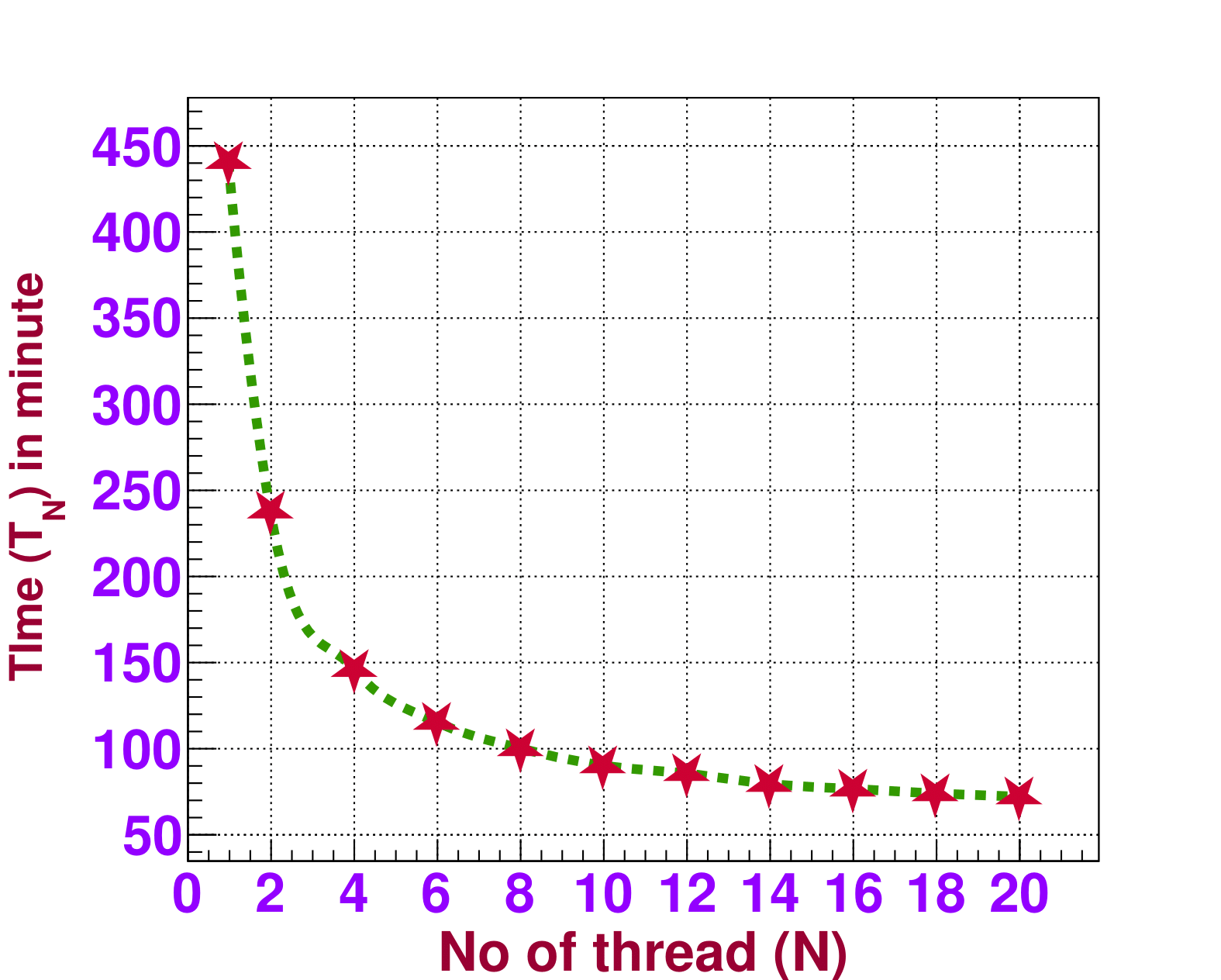}}\subfloat[\label{fig:speedUp_wsp}]{\includegraphics[width=.5\linewidth]{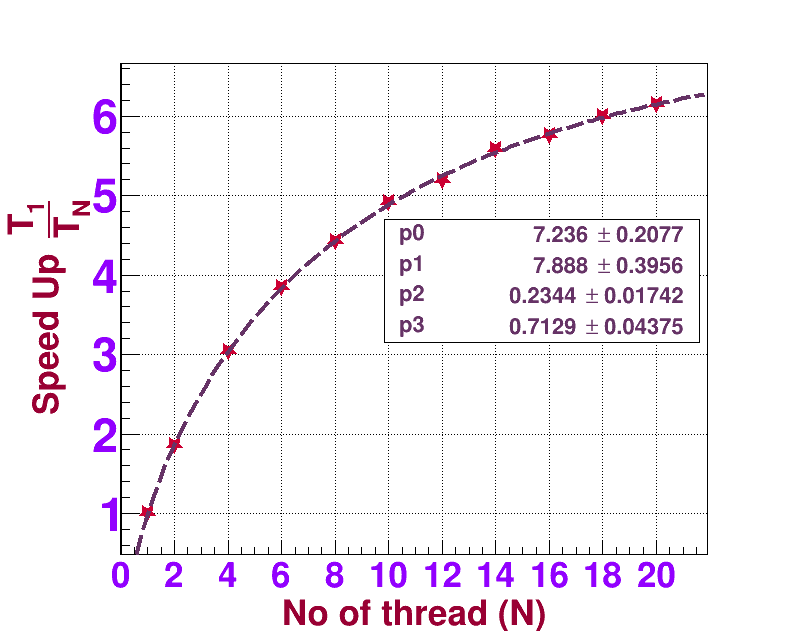}}
		
		\caption{ Variation of execution time to complete $10^4$ avalanches with number of threads (a) without space-charge effect and (c) with space-charge effect; speed up performance with number of threads (b) without space-charge effect, (d) with space-charge effect.}
		
	\end{figure}

	A single primary electron is placed near the negative electrode to generate avalanches.
	A set of $10^4$ avalanches has been generated repeatedly by varying the number of thread values ($N$) in OpenMP.
	Let $T_N$ represent the time taken to generate $10^4$ events with a total $N$ number of threads.
	The variation in total time ($T_N$) with and without space-charge effect as a function of $N$ threads has been shown in Figs. \ref{fig:timePer_nosp} and \ref{fig:timePer_wsp}, respectively.
	It is clear from the preliminary figures that in both cases, with and without the space-charge effect, $T_N$ decreases with the increase in the total number of threads $N$.
	The speed-up performance of the code can be calculated by taking the ratios between sequential time ($T_1$) with $T_N$. \textcolor{black}{Therefore, speed-up factor $v_p$ can be written as}:
	  	\begin{equation}
	  	v_p=\frac{T_1}{T_N}.
	  \end{equation}
    The speed-up performance $v_p$ as a function of number of threads ($N$) has been shown in Figs. \ref{fig:speedUp_nosp} and \ref{fig:speedUp_wsp} with and without space-charge effect, respectively. 
    The $v_p$ increases with the number of threads ($N$).
    The data points of both Figs.  \ref{fig:speedUp_nosp} and \ref{fig:speedUp_wsp} is fitted with a non linear function: 
	\begin{equation}\label{eqn:speed_up}
	f(N)=p_0-p_1\,e^{-(p_2\,N^{p_3})},\,(N\ge 1)
	\end{equation}
	 where the fit parameters are shown in the same figures. From equation \ref{eqn:speed_up}, it can be said that for large $N$, the contribution of the second term is significantly less; hence the parameter $p_0$ signifies the maximum speed-up which can be achieved.
	 Therefore, based on parameter $p_0$, the maximum speed up of 5.46 and 7.2 have been observed without and with space-charge effect, respectively.
	 \par The saturation in speed-up can be explained by overhead issues. 
	 The overhead arises during the communication and synchronization between threads.
	 More details about the overhead in OpenMP can be found in~\cite{openmp_overhead1,openmp_overhead2}.
	 It is also noted that the time plotted in Figs. \ref{fig:timePer_nosp} and \ref{fig:timePer_wsp} is the total time to complete all avalanches, which is the combination of time of parallel and non-parallel computations.
	 Therefore, $T_N$ is not the time purely taken by $N$ threads; instead, it is the sum of time by $N$ threads for parallel computations and a single thread for non-parallel or sequential computations. 
	
	\par
\begin{figure}
	\center\subfloat[\label{fig:Elc_gain_nosp}]{\includegraphics[width=.5\linewidth]{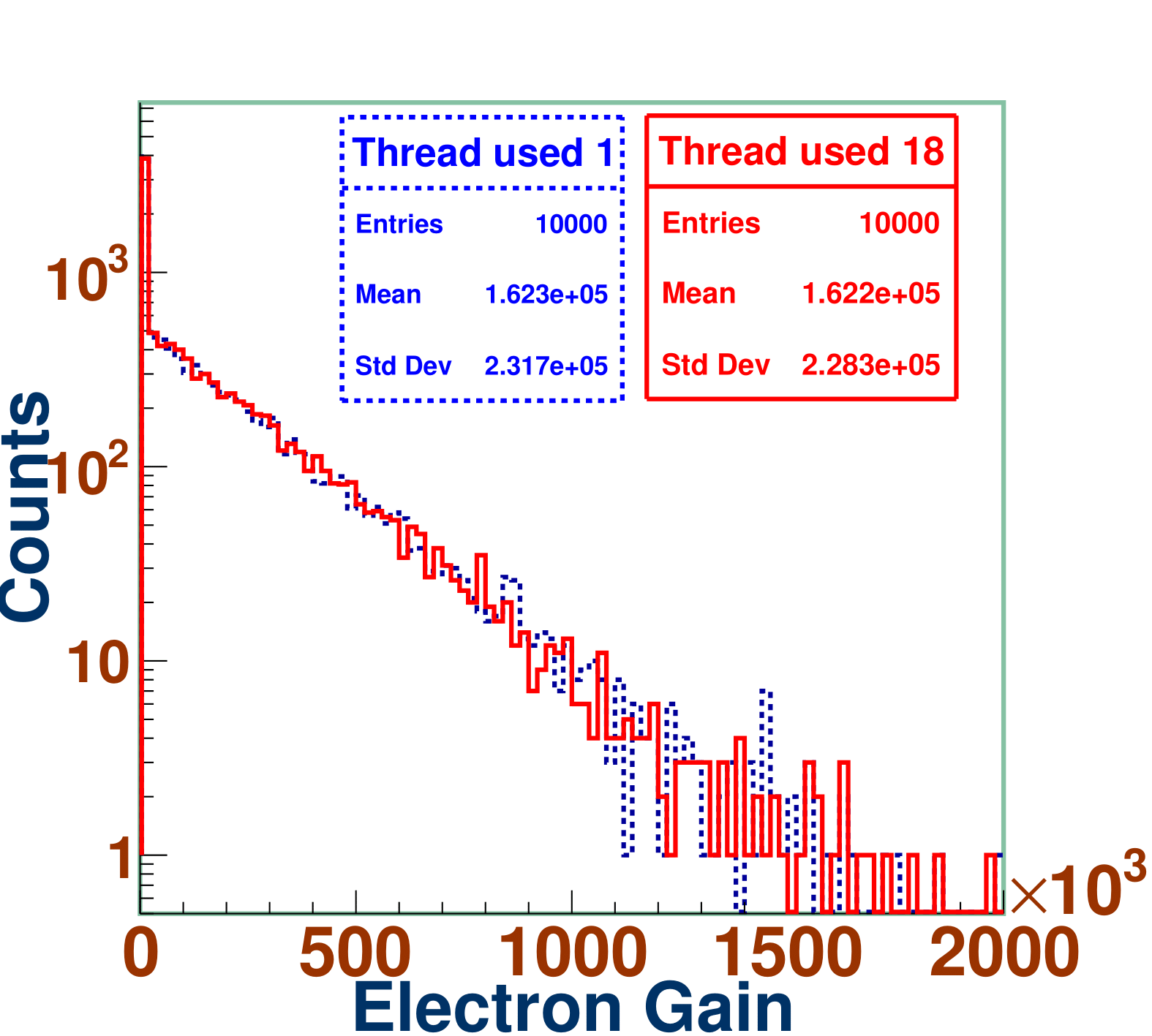}}\subfloat[\label{fig:Elc_gain_wsp}]{\includegraphics[width=.5\linewidth]{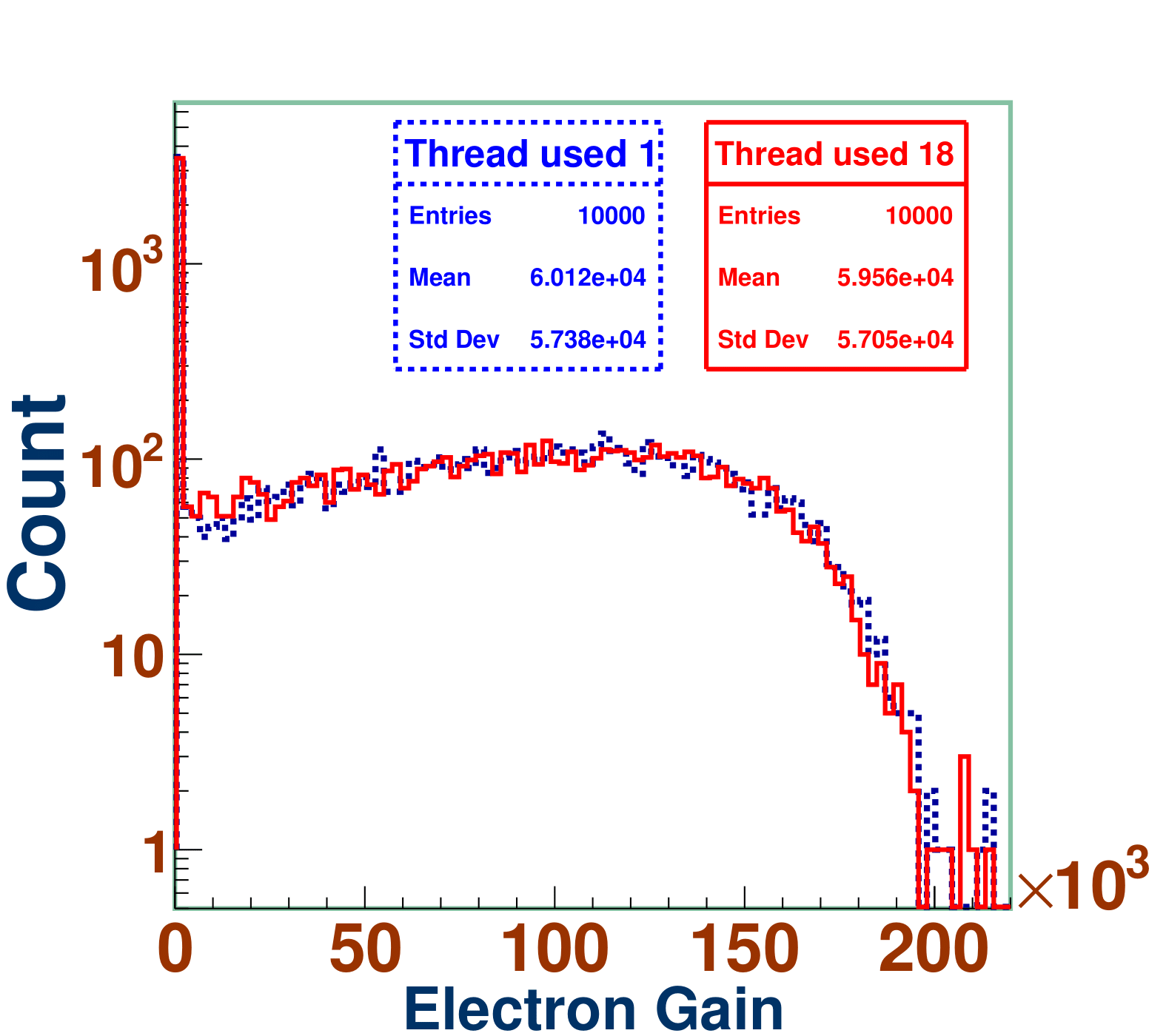}}
	\caption{Distribution of electron gain of $10^4$ avalanches (a) without space-charge effect (b) with space-charge effect.\label{fig:gain_comp}}
	
      \end{figure}

	In Figs. \ref{fig:Elc_gain_nosp} and \ref{fig:Elc_gain_wsp}, the distributions of electron gain of $10^4$ avalanches have been compared with and without space-charge effect, respectively.
	As a result, it is found \textcolor{black}{that the mean gain is reduced by order of 10 when the space-charge effect is considered, and the shape of the distribution is also modified significantly}.
	 Again, it is expected that the average electron gain should not vary much with the number of threads ($N$).
	 In the same Figs. \ref{fig:Elc_gain_nosp} and \ref{fig:Elc_gain_wsp}, the gain calculations have been carried out with single thread and a total of 18 threads, where it can be seen that they are matched within some statistical uncertainties.
	
	\section{Induced-charge distribution of avalanches}
	\label{sec:7_neBEM}
    Charge can be induced on the readout electrodes due to the movement of the electrons and ions.
	    The induced charge $Q^{ind}$ due to $N_{av}$ number of $q$ point charge moving inside the RPC can be calculated by using Ramo's equation as follows \cite{ShockleyRamo}:
     \begin{equation}\label{eqn:induced_charge}
    Q^{ind}=\int_{0}^{t} dt \sum_{n=0}^{N_{av}}\, \,q\,({\phi}_w^n({r_f}(t))-{\phi}_w^n({r_i}(t))),
    \end{equation}
    where ${\phi}_w^n({r_f}(t))$ and $\,{\phi}_w^n({r_i}(t)$ are the weighting potential at initial ($r_i$) and final position ($r_f$) of the step calculated by using neBEM.
   
      	\par
\begin{figure}[H]
		\center{\includegraphics[width=.7\linewidth]{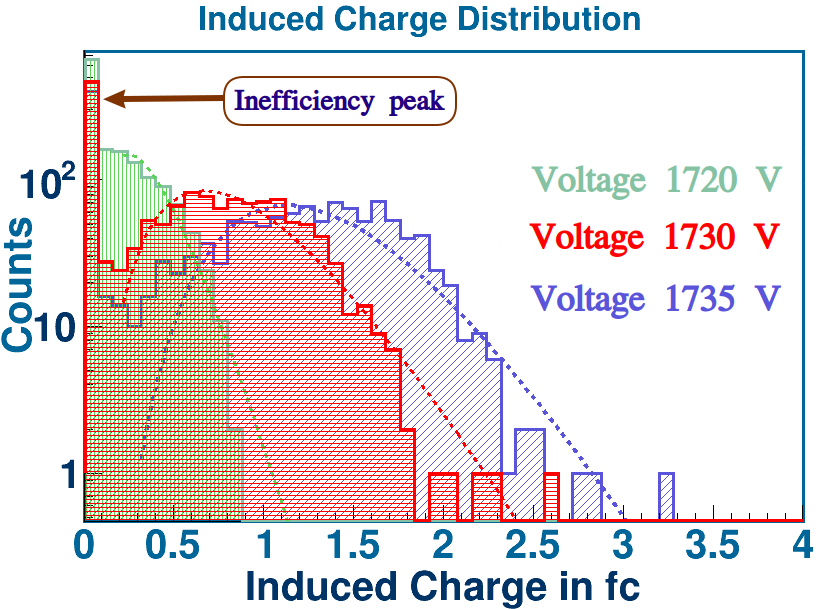}}
		\caption{Comparison between induced-charge distributions corresponding to three applied voltages 1720 V, 1730 V and 1735 V.\label{fig:induced_charge}}
		
	\end{figure}

      	The induced-charge distributions of a timing RPC for three different applied voltages, 1720 V, 1730 V, and 1735 V, are shown in Fig. \ref{fig:induced_charge}.
      	The RPC's geometry and other configurations remain the same as in Section \ref{sec:6_speedUp}.
      	The three distributions of Fig. \ref{fig:induced_charge} have been fitted with the Polya function given as follows \cite{polya_charge,KOBAYASHI2006136} :
		\begin{equation}
		f(Q^{ind})=a(\frac{Q^{ind}\,b}{c})^{b-1}\,e^{-\frac{b}{c}Q^{ind}},
		\end{equation}
	where parameter $a$ is the scaling factor, $b$ is a free parameter determining the shape of the distribution, and $c$ is the mean charge.
	In all fitting processes with Polya function, the left inefficiency peak of the charge distributions of the Fig. \ref{fig:induced_charge} has not been considered.
	The fitted values of $a$, $b$, and $c$ for different voltages given in the Table \ref{tab:ind_charge}.
	From the values of fit parameters of Fig. \ref{fig:induced_charge}, it can be said that the value of the parameter $c$ or mean charge increases with the increase in applied voltage. 
	Moreover, the value of $b$ also shifted towards a higher value with the voltage increment, determining the broadness of the charge spectrum.
	From experimental results such as \cite{Fonte:491918, LippmanThesis}, the shifting of the mean and broadening of the shape of induced-charge distribution with increasing applied voltage are also observed.
	 \begin{table}
	 	\center%
	 	\begin{tabular}{|c|c|c|c|c|}
	 		\hline 
	 		Voltage (V) & a & b & c (fc)
	 		\tabularnewline
	 		\hline 
	 		\hline 
	 		1720 & $\begin{array}{c}
	 		139.7\\
	 		\pm\\
	 		68.7
	 		\end{array}$ & $\begin{array}{c}
	 		3.8\\
	 		\pm\\
	 		0.5
	 		\end{array}$ & $\begin{array}{c}
	 		0.34\\
	 		\pm\\
	 		0.01
	 		\end{array}$\tabularnewline
	 		\hline 
	 		1730 & $\begin{array}{c}
	 		19.8\\
	 		\pm\\
	 		7.9
	 		\end{array}$ & $\begin{array}{c}
	 		4.9\\
	 		\pm\\
	 		0.3
	 		\end{array}$ & $\begin{array}{c}
	 		0.8\\
	 		\pm\\
	 		0.01
	 		\end{array}$\tabularnewline
	 		\hline 
	 		1735 & $\begin{array}{c}
	 		0.06\\
	 		\pm\\
	 		0.07
	 		\end{array}$ & $\begin{array}{c}
	 		8.2\\
	 		\pm\\
	 		0.7
	 		\end{array}$ & $\begin{array}{c}
	 		1.3\\
	 		\pm\\
	 		0.02
	 		\end{array}$ \tabularnewline
	 		\hline 
	 	\end{tabular}
	 	
	 	\caption{Fit parameters of induced-charge distribution of Fig. \ref{fig:induced_charge}\label{tab:ind_charge}.}
	 	
	 \end{table} 
	 
	\section{Summary}
	\label{sec:8_summary}
	
	In this paper, we have discussed the method of parallel generation of uncorrelated and uniform random numbers.
	The problem of correlation between the random numbers generated by multiple threads has been successfully solved using different seed values and objects of TRandom3.
	\par
	The steps of implementing the dynamic space-charge field inside Garfield++ have been discussed.
	In this context, the detailed modeling of a saturated avalanche has been examined to understand the space-charge effect.
	The changing of the dynamic space-charge field and the evolution of the 3D shape of the avalanche with time have also been shown.
	As a result, we have found that at the time of saturation, the rate of attachment is dominant in the low-field regions, while in the high-field regions the rate of ionization is dominant.
	Hence, overall, the number of charges remains the same, which brings this saturation.   
	 \par
	\textcolor{black}{The influence of negative ions has also been carefully examined. In both gas mixtures containing (a) Ar and CO$_2$ and (b) C$_2$H$_2$F$_4$, i-C$_4$H$_10$, and SF$_6$, electron gain is significantly reduced due to the presence of negative ions. The simulation shows instability in the absence of negative ions for the gas mixture containing SF$_6$, which is resolved when they are included in the simulation. } 
	
	\par
	\textcolor{black}{Parallelization has been acheived by implementing multithreading techniques in Garfield++.
	With OpenMP implemented in neBEM and the avalanche generation process has made it possible to achieve approximately 5.4 times and 7.2 times speed up without and with the space-charge effect, respectively.}
	\par
	Finally, the induced-charge distribution for a timing RPC has been calculated for three different voltages.
	The shape of the spectra of the induced-charge distribution predicted by the present simulations are similar to those observed in other experimental results \cite{Fonte:491918}.
	
      \par
      In future, we plan to implement a photon transport model in our code to study more effectively the avalanche-to-streamer transition.
      Moreover, the study of other gaseous detectors will also be possible with the model described in the present paper.

\section*{Acknowledgement}
All authors are grateful to the INO collaboration, the HEP experiment division of VECC, Applied Nuclear Physics Division of SINP and Physics Department, School of Basic and Applied Sciences, Adamas University for providing the resources and help. The authors thank the reviewer for constructive and meticulous review that helped improvement in both content and form of the manuscript.

\par
	
	
	\bibliography{pGarfield.bib}

\begin{thebibliography}{10}
\expandafter\ifx\csname url\endcsname\relax
  \def\url#1{\texttt{#1}}\fi
\expandafter\ifx\csname urlprefix\endcsname\relax\def\urlprefix{URL }\fi
\expandafter\ifx\csname href\endcsname\relax
  \def\href#1#2{#2} \def\path#1{#1}\fi

\bibitem{SAULI20162}
F.~Sauli, The gas electron multiplier ({{\color{black}GEM}}): Operating
  principles and applications, Nuclear Instruments and Methods in Physics
  Research Section A: Accelerators, Spectrometers, Detectors and Associated
  Equipment 805 (2016) 2--24, special Issue in memory of Glenn F. Knoll.
\newblock \href {https://doi.org/https://doi.org/10.1016/j.nima.2015.07.060}
  {\path{doi:https://doi.org/10.1016/j.nima.2015.07.060}}.

\bibitem{cardeli-1}
R.~Santonico, R.~Cardarelli, Development of resistive plate counters, Nuclear
  Instruments and Methods in Physics Research 187~(2) (1981) 377--380.
\newblock \href {https://doi.org/https://doi.org/10.1016/0029-554X(81)90363-3}
  {\path{doi:https://doi.org/10.1016/0029-554X(81)90363-3}}.

\bibitem{cardeli-2}
{\color{black}R. Cardarelli and R. Santonico and A.Di Biagio and A. Lucci},
  Progress in resistive plate counters, Nuclear Instruments and Methods in
  Physics Research Section A: Accelerators, Spectrometers, Detectors and
  Associated Equipment 263~(1) (1988) 20 -- 25.
\newblock \href {https://doi.org/https://doi.org/10.1016/0168-9002(88)91011-X}
  {\path{doi:https://doi.org/10.1016/0168-9002(88)91011-X}}.

\bibitem{Hilke:2010zz}
H.~J. Hilke, {Time projection chambers}, Rept. Prog. Phys. 73 (2010) 116201.
\newblock \href {https://doi.org/10.1088/0034-4885/73/11/116201}
  {\path{doi:10.1088/0034-4885/73/11/116201}}.

\bibitem{Goswami_2017}
S.~Goswami, The status of {INO}, Journal of Physics: Conference Series 888
  (2017) 012025.
\newblock \href {https://doi.org/10.1088/1742-6596/888/1/012025}
  {\path{doi:10.1088/1742-6596/888/1/012025}}.

\bibitem{Kumari_2020}
{\color{black}P. Kumari et al.}, Improved-{RPC} for the {CMS} muon system
  upgrade for the {HL}-{LHC}, Journal of Instrumentation 15~(11) (2020)
  C11012--C11012.
\newblock \href {https://doi.org/10.1088/1748-0221/15/11/c11012}
  {\path{doi:10.1088/1748-0221/15/11/c11012}}.

\bibitem{Collaboration_2012}
{{\color{blue}F. Boss\`{u} et al.}}, {Performance of the {RPC}-based {ALICE}
  muon trigger system at the {LHC}}, {Journal of Instrumentation} {7}~({12})
  (2012) {T12002--T12002}.
\newblock \href {https://doi.org/{10.1088/1748-0221/7/12/t12002}}
  {\path{doi:{10.1088/1748-0221/7/12/t12002}}}.

\bibitem{MONDAL2021166042}
M.~Mondal, T.~Dey, S.~Chattopadhyay, J.~Saini, Z.~Ahammed, {Performance of a
  prototype bakelite {\color{black}RPC} at {\color{black}GIF++} using
  self-triggered electronics for the {\color{black}CBM} experiment at
  {\color{black}FAIR}}, Nuclear Instruments and Methods in Physics Research
  Section A: Accelerators, Spectrometers, Detectors and Associated Equipment
  (2021) 166042\href {https://doi.org/{
  https://doi.org/10.1016/j.nima.2021.166042}} {\path{doi:{
  https://doi.org/10.1016/j.nima.2021.166042}}}.

\bibitem{MOSHAII2012S168}
A.~Moshaii, L.~K. Khorashad, M.~Eskandari, S.~Hosseini, {\color{black}RPC
  simulation in avalanche and streamer modes using transport equations for
  electrons and ions}, Nuclear Instruments and Methods in Physics Research
  Section A: Accelerators, Spectrometers, Detectors and Associated Equipment
  661 (2012) S168--S171, x. Workshop on Resistive Plate Chambers and Related
  Detectors (RPC 2010).
\newblock \href {https://doi.org/https://doi.org/10.1016/j.nima.2010.09.133}
  {\path{doi:https://doi.org/10.1016/j.nima.2010.09.133}}.

\bibitem{Datta_2020}
J.~Datta, S.~Tripathy, N.~Majumdar, S.~Mukhopadhyay,
  \href{https://dx.doi.org/10.1088/1748-0221/15/12/C12006}{Study of streamer
  development in resistive plate chamber}, Journal of Instrumentation 15~(12)
  (2020) C12006.
\newblock \href {https://doi.org/10.1088/1748-0221/15/12/C12006}
  {\path{doi:10.1088/1748-0221/15/12/C12006}}.
\newline\urlprefix\url{https://dx.doi.org/10.1088/1748-0221/15/12/C12006}

\bibitem{PFonte_2013}
P.~Fonte, \href{https://dx.doi.org/10.1088/1748-0221/8/11/P11001}{Survey of
  physical modelling in resistive plate chambers}, Journal of Instrumentation
  8~(11) (2013) P11001.
\newblock \href {https://doi.org/10.1088/1748-0221/8/11/P11001}
  {\path{doi:10.1088/1748-0221/8/11/P11001}}.
\newline\urlprefix\url{https://dx.doi.org/10.1088/1748-0221/8/11/P11001}

\bibitem{Rout_2021}
P.~Rout, J.~Datta, P.~Roy, P.~Bhattacharya, S.~Mukhopadhyay, N.~Majumdar,
  S.~Sarkar, \href{https://dx.doi.org/10.1088/1748-0221/16/02/P02018}{{Fast
  simulation of avalanche and streamer in {\color{black}GEM} detector using
  hydrodynamic approach}}, Journal of Instrumentation 16~(02) (2021) P02018.
\newblock \href {https://doi.org/10.1088/1748-0221/16/02/P02018}
  {\path{doi:10.1088/1748-0221/16/02/P02018}}.
\newline\urlprefix\url{https://dx.doi.org/10.1088/1748-0221/16/02/P02018}

\bibitem{LippmanThesis}
C.~Lippmann,
  \href{https://cds.cern.ch/record/1303626/files/CERN-THESIS-2003-035.pdf}{Detector
  physics of resistive plate chambers (cern-thesis-2003-035)} (2003).
\newline\urlprefix\url{https://cds.cern.ch/record/1303626/files/CERN-THESIS-2003-035.pdf}

\bibitem{Lippmann_1}
C.~Lippmann, W.~Riegler, {Space charge effects in resistive plate chambers},
  Nucl. Instrum. Meth. A 517 (2004) 54--76.
\newblock \href {https://doi.org/10.1016/j.nima.2003.08.174}
  {\path{doi:10.1016/j.nima.2003.08.174}}.

\bibitem{Lippmann:2003ar}
C.~Lippmann, W.~Riegler, B.~Schnizer, {Space charge effects and induced signals
  in resistive plate chambers}, Nucl. Instrum. Meth. A 508 (2003) 19--22.
\newblock \href {https://doi.org/10.1016/S0168-9002(03)01270-1}
  {\path{doi:10.1016/S0168-9002(03)01270-1}}.

\bibitem{Garfield}
H.~Schindler, {Garfield++ User$'$s Guide},
  \href{https://garfieldpp.web.cern.ch/garfieldpp/documentation/UserGuide.pdf}{\url{https://garfieldpp.web.cern.ch/garfieldpp}}
  (April 2020).

\bibitem{Veenhof:1993hz}
R.~Veenhof, {Garfield, a drift chamber simulation program}, Conf. Proc. C
  9306149 (1993) 66--71.

\bibitem{VEENHOF1998726}
R.~Veenhof, Garfield, recent developments, Nuclear Instruments and Methods in
  Physics Research Section A: Accelerators, Spectrometers, Detectors and
  Associated Equipment 419~(2) (1998) 726--730.
\newblock \href {https://doi.org/https://doi.org/10.1016/S0168-9002(98)00851-1}
  {\path{doi:https://doi.org/10.1016/S0168-9002(98)00851-1}}.

\bibitem{comsol}
\href{https://www.comsol.co.in/}{https://www.comsol.co.in/}.

\bibitem{MAJUMDAR2008346}
N.~Majumdar, S.~Mukhopadhyay, S.~Bhattacharya, {Computation of
  {\color{black}3D} electrostatic weighting field in Resistive Plate Chambers},
  Nuclear Instruments and Methods in Physics Research Section A: Accelerators,
  Spectrometers, Detectors and Associated Equipment 595~(2) (2008) 346--352.
\newblock \href {https://doi.org/https://doi.org/10.1016/j.nima.2008.07.033}
  {\path{doi:https://doi.org/10.1016/j.nima.2008.07.033}}.

\bibitem{MAJUMDAR2009719}
N.~Majumdar, S.~Mukhopadhyay, S.~Bhattacharya, Three-dimensional electrostatic
  field simulation of a resistive plate chamber, Nuclear Instruments and
  Methods in Physics Research Section A: Accelerators, Spectrometers, Detectors
  and Associated Equipment 602~(3) (2009) 719--722, proceedings of the 9th
  International Workshop on Resistive Plate Chambers and Related Detectors.
\newblock \href {https://doi.org/https://doi.org/10.1016/j.nima.2008.12.098}
  {\path{doi:https://doi.org/10.1016/j.nima.2008.12.098}}.

\bibitem{OpenMp}
\href{https://www.openmp.org/}{https://www.openmp.org/}.

\bibitem{Dey_2020}
T.~Dey, S.~Mukhopadhyay, S.~Chattopadhyay, J.~Sadukhan, Numerical study of
  space charge electric field inside resistive plate chamber, Journal of
  Instrumentation 15~(11) (2020) C11005--C11005.
\newblock \href {https://doi.org/10.1088/1748-0221/15/11/c11005}
  {\path{doi:10.1088/1748-0221/15/11/c11005}}.

\bibitem{Dey_2022}
T.~Dey, S.~Mukhopadhyay, S.~Chattopadhyay, Numerical study of effects of
  electrode parameters and image charge on the electric field configuration of
  {RPCs}, Journal of Instrumentation 17~(04) (2022) P04015.
\newblock \href {https://doi.org/10.1088/1748-0221/17/04/p04015}
  {\path{doi:10.1088/1748-0221/17/04/p04015}}.

\bibitem{pGarfield1}
{Sheharyar, A.}, {Bouhali, O.}, {Castaneda, A.}, Speeding up and parallelizing
  the garfield++, EPJ Web Conf. 174 (2018) 06004.
\newblock \href {https://doi.org/10.1051/epjconf/201817406004}
  {\path{doi:10.1051/epjconf/201817406004}}.

\bibitem{BOUHALI201892}
O.~Bouhali, A.~Sheharyar, T.~Mohamed, Accelerating avalanche simulation in gas
  based charged particle detectors, Nuclear Instruments and Methods in Physics
  Research Section A: Accelerators, Spectrometers, Detectors and Associated
  Equipment 901 (2018) 92--98.
\newblock \href {https://doi.org/https://doi.org/10.1016/j.nima.2018.05.061}
  {\path{doi:https://doi.org/10.1016/j.nima.2018.05.061}}.

\bibitem{root-cern}
{{\color{black}ROOT} Data analysis framework User$'$s Guide},
  \href{https://root.cern.ch/root/htmldoc/guides/users-guide/ROOTUsersGuideA4.pdf}{\url{https://root.cern.ch/root/htmldoc/guides/users-guide/ROOTUsersGuideA4.pdf}}
  (May 2018).

\bibitem{BrunRoot}
R.~Brun, F.~Rademakers, {ROOT: An object oriented data analysis framework},
  Nucl. Instrum. Meth. A 389 (1997) 81--86.
\newblock \href {https://doi.org/10.1016/S0168-9002(97)00048-X}
  {\path{doi:10.1016/S0168-9002(97)00048-X}}.

\bibitem{MersenneTwister}
M.~M.~T. Nishimura, Mersenne twister: {A} 623-dimensionally equidistributed
  uniform pseudo-random number generator (Jan 1998).
\newblock \href {https://doi.org/https://doi.org/10.1145/272991.272995}
  {\path{doi:https://doi.org/10.1145/272991.272995}}.

\bibitem{SMIRNOV2005474}
I.~Smirnov, Modeling of ionization produced by fast charged particles in gases,
  Nuclear Instruments and Methods in Physics Research Section A: Accelerators,
  Spectrometers, Detectors and Associated Equipment 554~(1) (2005) 474--493.
\newblock \href {https://doi.org/https://doi.org/10.1016/j.nima.2005.08.064}
  {\path{doi:https://doi.org/10.1016/j.nima.2005.08.064}}.

\bibitem{BIAGI1989716}
A multiterm boltzmann analysis of drift velocity, diffusion, gain and
  magnetic-field effects in argon-methane-water-vapour mixtures, Nuclear
  Instruments and Methods in Physics Research Section A: Accelerators,
  Spectrometers, Detectors and Associated Equipment 283~(3) (1989) 716--722.
\newblock \href {https://doi.org/https://doi.org/10.1016/0168-9002(89)91446-0}
  {\path{doi:https://doi.org/10.1016/0168-9002(89)91446-0}}.

\bibitem{BIAGI1999234}
S.~Biagi, Monte carlo simulation of electron drift and diffusion in counting
  gases under the influence of electric and magnetic fields, Nuclear
  Instruments and Methods in Physics Research Section A: Accelerators,
  Spectrometers, Detectors and Associated Equipment 421~(1) (1999) 234--240.
\newblock \href {https://doi.org/https://doi.org/10.1016/S0168-9002(98)01233-9}
  {\path{doi:https://doi.org/10.1016/S0168-9002(98)01233-9}}.

\bibitem{Furry}
W.~H. Furry, \href{https://link.aps.org/doi/10.1103/PhysRev.52.569}{On
  fluctuation phenomena in the passage of high energy electrons through lead},
  Phys. Rev. 52 (1937) 569--581.
\newblock \href {https://doi.org/10.1103/PhysRev.52.569}
  {\path{doi:10.1103/PhysRev.52.569}}.
\newline\urlprefix\url{https://link.aps.org/doi/10.1103/PhysRev.52.569}

\bibitem{Schindler:1500583}
H.~Schindler, \href{https://cds.cern.ch/record/1500583}{{Microscopic Simulation
  of Particle Detectors}}, presented 13 Dec 2012 (2012).
\newline\urlprefix\url{https://cds.cern.ch/record/1500583}

\bibitem{openmp_overhead1}
J.~M. Bull, Measuring synchronisation and scheduling overheads in openmp, in:
  In Proceedings of First European Workshop on OpenMP, 1999, pp. 99--105.

\bibitem{openmp_overhead2}
{{\color{blue}M. K. Bane and G. D. Riley}}, {Extended Overhead Analysis for
  {\color{black}OpenMP}}, Springer Berlin Heidelberg, Berlin, Heidelberg, 2002,
  pp. 162--166.
\newblock \href {https://doi.org/https://doi.org/10.1007/3-540-45706-2\_20}
  {\path{doi:https://doi.org/10.1007/3-540-45706-2\_20}}.

\bibitem{ShockleyRamo}
Z.~He, Review of the shockley ramo theorem and its application in semiconductor
  gamma-ray detectors, Nuclear Instruments and Methods in Physics Research
  Section A: Accelerators, Spectrometers, Detectors and Associated Equipment
  463~(1) (2001) 250--267.
\newblock \href {https://doi.org/https://doi.org/10.1016/S0168-9002(01)00223-6}
  {\path{doi:https://doi.org/10.1016/S0168-9002(01)00223-6}}.

\bibitem{polya_charge}
{{\color{black}M Abbrescia et al.}}, The simulation of resistive plate chambers
  in avalanche mode: charge spectra and efficiency, Nuclear Instruments and
  Methods in Physics Research Section A: Accelerators, Spectrometers, Detectors
  and Associated Equipment 431~(3) (1999) 413--427.
\newblock \href {https://doi.org/https://doi.org/10.1016/S0168-9002(99)00374-5}
  {\path{doi:https://doi.org/10.1016/S0168-9002(99)00374-5}}.

\bibitem{KOBAYASHI2006136}
M.~Kobayashi, {An estimation of the effective number of electrons contributing
  to the coordinate measurement with a {\color{black}TPC}}, Nuclear Instruments
  and Methods in Physics Research Section A: Accelerators, Spectrometers,
  Detectors and Associated Equipment 562~(1) (2006) 136--140.
\newblock \href {https://doi.org/https://doi.org/10.1016/j.nima.2006.03.001}
  {\path{doi:https://doi.org/10.1016/j.nima.2006.03.001}}.

\bibitem{Fonte:491918}
P.~Fonte, V.~Peskov, {High-resolution {\color{black}TOF} with
  {\color{black}RPCs}} (2001) 6 p\href
  {https://doi.org/10.1016/S0168-9002(01)01914-3}
  {\path{doi:10.1016/S0168-9002(01)01914-3}}.

\end{thebibliography}
	\bibliographystyle{elsarticle-harv.bst}

	
	
	
\end{document}